%% file: EPIC_216_EB_FINAL_arXiv.tex
\UseRawInputEncoding
\documentclass[a4paper,fleqn,usenatbib]{mnras}
\usepackage[T1]{fontenc}
\usepackage{ae,aecompl}
\usepackage{hyperref}

\usepackage{graphicx}	
\usepackage{amsmath}	
\usepackage{amssymb}	

\input{defs}  

\title[EPIC\,216747137: a new sdOB+dM HW\,Vir eclipsing binary]
{EPIC\,216747137: a new HW\,Vir eclipsing binary with a massive sdOB 
primary and a low-mass M-dwarf companion} 
\author[]{
R. Silvotti,$^{1}$\thanks{E-mail: roberto.silvotti@inaf.it}
V. Schaffenroth,$^{2}$
U. Heber,$^{3}$
R.\,H. \O stensen,$^{4}$
J.\,H. Telting,$^{5}$
J. Vos,$^{2}$
\newauthor
\hspace{0.0mm}
D. Kilkenny,$^{6}$
L. Mancini,$^{7,8,1}$
S. Ciceri,$^{9}$
A. Irrgang,$^{3}$
H. Drechsel\,$^{3}$
\vspace{2mm}
\\
$^{1}$INAF-Osservatorio Astrofisico di Torino, Strada dell'Osservatorio 20, 
10025 Pino Torinese, Italy\\
$^{2}$University of Potsdam, Institute of Physics and Astronomy,
Karl-Liebknecht-Str. 24-25, 14476 Potsdam, Germany\\
$^{3}$Dr. Remeis-Sternwarte \& ECAP, Astronomical Institute, 
University of Erlangen-N\"{u}rnberg, Sternwartstr. 7,
96049 Bamberg, Germany\\
$^{4}$Department of Physics, Astronomy and Materials Science, Missouri State 
University, 901 S. National, Springfield, MO 65897, USA\\
$^{5}$Nordic Optical Telescope, Rambla José Ana Fernández Pérez 7, 
38711 Breña Baja, Spain\\
$^{6}$Department of Physics \& Astronomy, University of the Western Cape, 
Private Bag X17, Bellville 7535, South Africa\\
$^{7}$Department of Physics, University of Rome ``Tor Vergata'', Via della 
Ricerca Scientifica 1, 00133 Roma, Italy\\
$^{8}$Max Planck Institute for Astronomy, K\"{o}nigstuhl 17,
69117 Heidelberg, Germany\\
$^{9}$simona.ciceri@gmail.com (Last Institution: Department of Astronomy, Stockholm University, Stockholm, Sweden)\\
}

\date{Accepted 2020 October 18. Received 2020 October 16; in original form 2020 October 1}

\pubyear{2020}

\begin{document}
\label{firstpage}
\pagerange{\pageref{firstpage}--\pageref{lastpage}}
\maketitle

\begin{abstract}
%
%
EPIC\,216747137 is a new HW~Virginis system discovered by the $Kepler$ 
spacecraft during its $K2$ ``second life''.
Like the other HW~Vir systems, EPIC\,216747137 is a post-common-envelope 
eclipsing binary consisting of a hot subluminous star and a cool low-mass 
companion.
The short orbital period of 3.87 hours produces a strong reflection effect
from the secondary ($\sim$9\% in the R band).
Together with AA\,Dor and V1828\,Aql, EPIC\,216747137 belongs to a 
small subgroup of HW~Vir systems with a hot evolved sdOB primary.
We find the following atmospheric parameters for the hot component:
\teff\,=\,40400$\pm$1000~K, \logg\,=\,5.56$\pm$0.06,
log($N$(He)/$N$(H))\,=\,$-2.59\pm$0.05.
The sdOB rotational velocity v\,$\sin$\,i\,=\,51$\pm$10~km\,s$^{-1}$
implies that the stellar rotation is slower than the orbital revolution
and the system is not synchronized.
When we combine photometric and spectroscopic results with the Gaia parallax, 
the best solution for the system corresponds to a primary with a mass of about
0.62~\msun\ close to, and likely beyond, the central helium exhaustion,
while the cool M-dwarf companion has a mass of about 0.11~\msun.
\end{abstract}





\section{Introduction}

Post-common-envelope binaries (PCEBs) are crucial to study the poorly 
understood and short-lived common-envelope phase of stellar evolution.

Among PCEBs, HW~Virginis stars are a specific class of eclipsing binaries 
consisting of a hot subdwarf primary with an M-dwarf companion
(see \citealt{Heber_2016} for a recent review on hot subdwarf stars).
There are two subgroups of HW~Vir stars: those similar to the prototype, with 
a core-helium-burning sdB (subdwarf B) primary, located in the extreme
horizontal branch (EHB) part of the H-R diagram. And those like 
AA\,Dor, with a hotter and more evolved primary of sdOB spectral class, beyond
the central helium exhaustion.

The possibility of measuring accurate dynamical masses in HW\,Virginis systems
is important to shed light on the formation mechanism of hot subdwarfs.
These stars are characterized by very thin hydrogen envelopes and masses 
close to the canonical mass of 0.47~\msun.

To form such an object, the hydrogen envelope of the red giant progenitor 
must be removed almost completely.
\citet[see also \citealt{2012ApJ...746..186C}]{Han_2002,Han_2003} describe 
three main binary evolution scenarios 
to form an sdB star: (i) common envelope (CE) ejection, (ii) stable 
Roche-lobe overflow (RLOF), and (iii) the merger of two He white dwarfs.
The latter scenario may contribute only for a very small fraction of sdBs 
given that the high masses and high rotation rates foreseen are not 
supported by the observations (\citealt{Fontaine_2012};
\citealt{Charpinet_2018,Reed_2018}).
Since $\sim$50\% of the non-composite-spectrum hot subdwarfs
are members of short-period binaries 
with orbital periods between 0.027 and $\sim$30 days 
\citep{Maxted_2001,Napiwotzki_2004,Kupfer_2015,Kupfer_2020},
mostly with a white dwarf (WD) or an M-type main-sequence (MS) 
companion, CE ejection triggered by a close companion is generally regarded as
the main formation channel.
As far as the RLOF scenario is concerned, an important recent work by 
\citet{Pelisoli_2020}
%
shows that almost all the wide binaries with K to F-type MS companions that 
they analysed show evidence of previous interaction,
confirming that the RLOF is another efficient way to form
$\sim$30-40\% of hot subdwarfs, and suggesting that binary interaction 
may {\it always} be required to form a hot subdwarf star.
Indeed, \citet{2003AJ....126.1455S} found that $\sim$40\% of hot subdwarfs 
have colors consistent with the presence of an unresolved late-type companion 
in a magnitude-limited sample (or $\sim$30\% in a volume-limited sample).
Putting these numbers together, we can estimate that $\sim$35\%
of hot subdwarfs are in close binaries with M-dwarf or WD companions, while
$\sim$30\% are in wide binaries with F/G/K companions.

However, the remaining fraction of $\sim$35\% consists of apparently single
hot subdwarfs.
For them, different formation mechanisms have been invoked, including
the merger of a He white dwarf with a low-mass hydrogen-burning star
\citep{2011ApJ...733L..42C}.
The presence of a substellar companion, difficult to detect, orbiting 
the sdB progenitor is another possibility \citep{Soker_1998, Han_2012},
only partially supported by the observations.

On the one hand, no planets transiting hot subdwarfs were found in a large 
survey with the Evryscope, capable of detecting planets with radii slightly 
smaller than Jupiter \citep{2020ApJ...890..126R}.
Neither were planetary transits of hot subdwarfs found up to now by the
Kepler/K2 or the TESS space missions.
Moreover, no significant radial velocity (RV) variations were found from 
high-accuracy Harps-N measurements of a small sample 
of 8 bright apparently single sdB stars \citep{Silvotti_2020},
excluding the presence of close substellar companions down to a few Jupiter 
masses and, for half of these stars, excluding also the presence 
of higher-mass companions in wide orbits.
These null results do not exclude that the planets were completely 
destroyed during the CE phase or that their envelope was removed 
leaving a very small and dense planetary core, difficult to detect
(see e.g. the controversial cases of KIC\,5807616 and KIC\,10001893, 
\citealt{2011Natur.480..496C, 2014A&A...570A.130S}).

On the other hand, there are at least three known HW-Vir systems with 
brown dwarf (BD) companions having masses between 0.04 and 0.07~\msun\
\citep{Geier_2011,Schaffenroth_2014,Schaffenroth_2015}, plus two more with 
masses close to the hydrogen-burning limit (\citealt{Schaffenroth_2019},
Fig.~14 and 15).
And there are a few controversial cases of planet detections through the
eclipse or pulsation timing method (see e.g. \citealt{2016IAUFM..29B.497B} 
and references therein).

Thanks to the high number of new HW-Vir systems discovered recently
from the light curves of the OGLE and ATLAS projects
\citep{Schaffenroth_2019}, and the new ones that are being discovered by the
TESS mission, the number of HW-Vir systems with substellar companions 
is likely to grow significantly in the short term.
With enough statistics it should be possible to determine the minimum
mass for a substellar companion to eject the red giant envelope
and survive engulfment. 
According to theory, it was thought that this limit could be 
near 10~\mjup\ \citep{Soker_1998,Han_2012}, but a recent article suggests 
that this mass limit could be higher, around 30-50~\mjup\ \citep{Kramer_2020}.
%
%


The system described in this paper, EPIC\,216747137 (alias UCAC2\,23234937), 
is a new HW-Vir binary
relatively bright (Gaia DR2 magnitude G\,=\,13.767$\pm$0.004), located about 
880\,pc from us (Gaia DR2 parallax of 1.14$\pm$0.06\,mas).
In the next sections we present the results of an analysis of both
photometric and spectroscopic data of EPIC\,216747137, that allow us
to infer the orbital parameters of the system and the main characteristics of 
the primary and secondary components.


\section{Time-series photometry} 

\subsection{$K2$ discovery}

\begin{figure}
\centering
\includegraphics[width=8.5cm,angle=0]{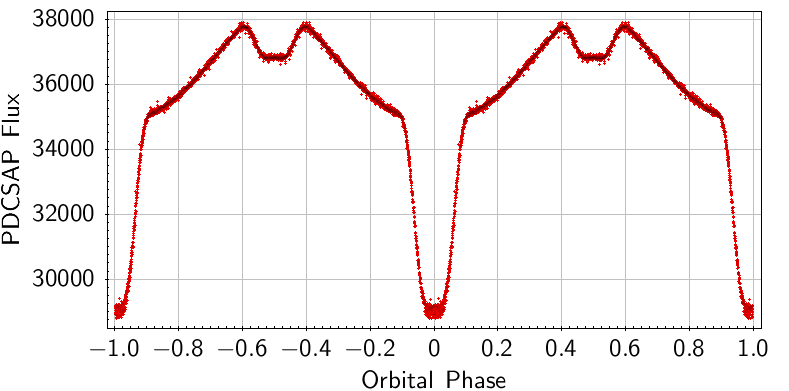}
\vspace{-3mm}
\caption{K2 light curve.}
\label{K2_lc}
\end{figure}

EPIC\,216747137 was observed by the {\it Kepler} space telescope during 
cycle 7
of its $K2$ secondary mission in long-cadence mode, with a sampling time of 
29.42\,minutes.
We downloaded the data from the ``Barbara A. Mikulski Archive for
Space Telescopes'' (MAST)\footnote{archive.stsci.edu} and we used the 
PDCSAP fluxes (PDC=Presearch Data Conditioning, SAP=Simple Aperture 
Photometry, see $K2$ documentation for more details).
After having removed some bad data point (those with SAP\_QUALITY flag 
different from zero or 2048 plus two outliers), the data set we used, as shown
in Fig.~\ref{K2_lc}, consists of 81.3\,days from BJD$_{TBD}$ 2457301.48620 to 
2457382.80453 (corresponding to 05/10/2015 -- 26/12/2015).

\subsection{SAAO BVR data}

EPIC\,216747137 was re-observed at the Sutherland site of the South African
Astronomical Observatory (SAAO) using the 1-m Elizabeth telescope
with the STE3 CCD photometer which has a read-out time of about 6\,s
(pre-binned 2$\times$2), small compared to the exposure times for filters
B (60~s), V (30~s) and R (30~s). Observations were made using each filter on 
a separate night (18, 17 and 19 May 2017, respectively) to maximise the 
resolution of the light curve.
Reduction of the CCD frames, magnitude extraction by profile-fitting, and
differential correction using several field stars were performed using
software written by Darragh O'Donoghue and partly based on the DoPHOT
program described by \citet{Schechter_1993}.

\begin{figure}
\centering
\includegraphics[width=8.5cm,angle=0]{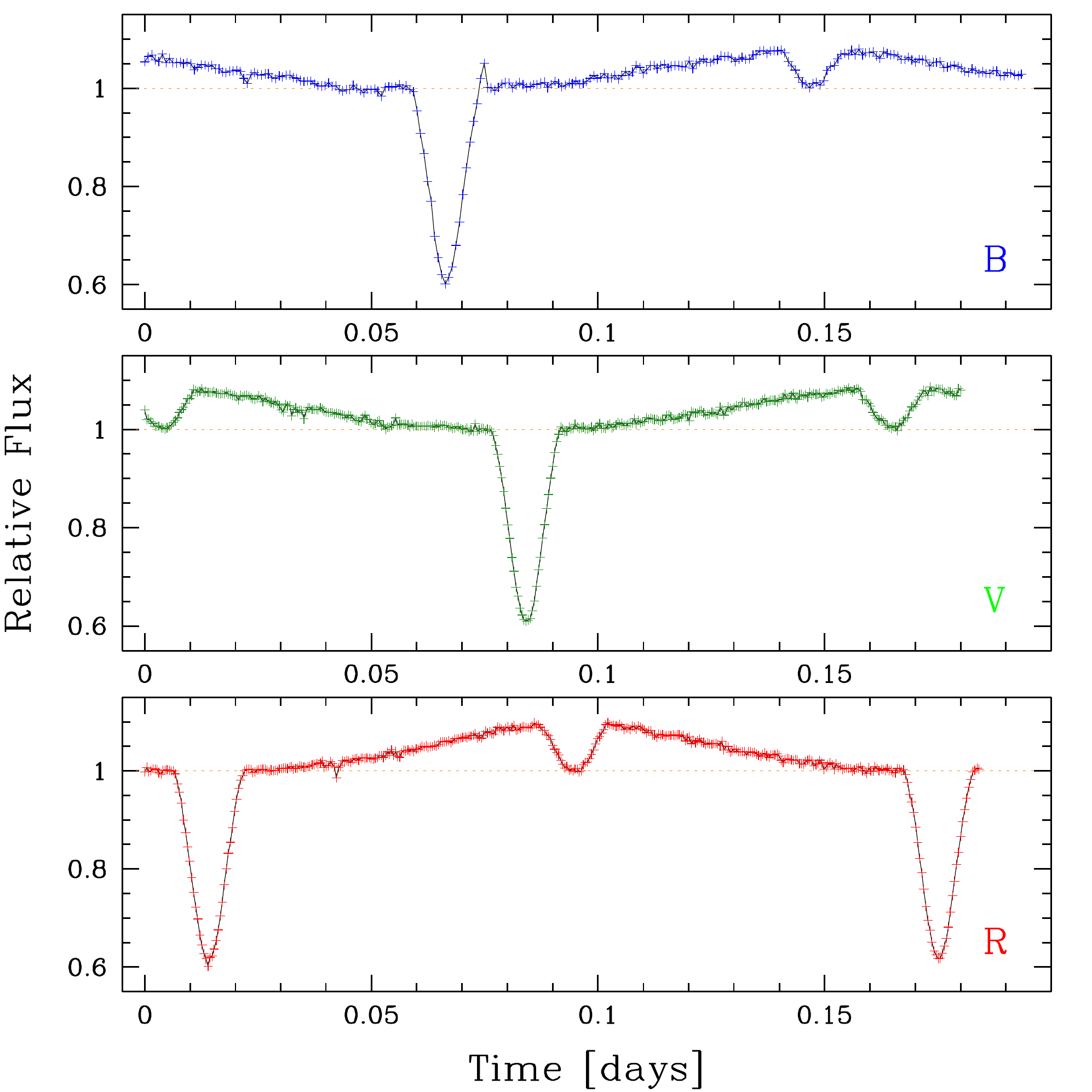}
\vspace{-3mm}
\caption{SAAO BVR light curves. The flux is normalized to the level just 
before and just after the primary eclipse.}
\label{SAAO_lc}
\end{figure}

The BVR light curves are shown in Fig.~\ref{SAAO_lc}.
Comparing Fig.~\ref{K2_lc} with Fig.~\ref{SAAO_lc}, we immediately note the
different shape and depth of the primary and secondary eclipses, due to the
smearing caused by the poor sampling rate of the $K2$ long-cadence data.
In Fig.~\ref{K2_lc} the primary and secondary eclipse have a depth of 
$\sim$17\% and less than 3\% respectively, while they are much deeper in the 
SAAO data ($\sim$39\% and $\sim$8\% in the R band).
For this reason, the $K2$ data was used only to improve the ephemeris, while 
the analysis of the light curve was performed using the ground-based 
photometry.

\section{Radial velocities}

EPIC\,216747137 was observed spectroscopically with various instruments.
As a first step, in order to measure the radial velocities (RVs) of the 
primary, 9 high-resolution spectra were obtained at different orbital phases 
in July and September 2016 using FEROS with the 2.2\,m MPG/ESO telescope 
at La Silla Observatory in Chile, with exposure times of
1800\,s\footnote{For the first two spectra we used 1000 and 1500\,s.}.
The FEROS spectra were reduced using CERES, a pipeline written for \'echelle 
spectrographs described in \citet{Brahm_2017}. The raw spectra were first 
corrected with calibration frames obtained in the afternoon or during 
twilight, and then calibrated in wavelength using a Th-Ar spectrum. 
The radial velocities of the sdOB star were measured from the He\,II line at 
4686\,\AA, while the Balmer lines were not used because they give more noisy 
results.
However, the results were quite poor due to the low SN ratio of the FEROS 
spectra.

For this reason, new observations were carried out as part of our $K2$ sdBV 
follow-up spectroscopic survey \citep{Telting_2014}.
We obtained 32 low-resolution spectra (R$\sim$2000, or 2.2~\AA) in two runs 
(22 spectra in July 2017, 10 spectra between March and August 2018)
at the 2.56\,m Nordic Optical Telescope (NOT, La Palma) using ALFOSC,
600\,s exposure times,
grism\#18, 0.5 arcsec slit, and CCD\#14, giving an approximate wavelength 
range 345-535\,nm.
The spectra were homogeneously reduced and analysed. Standard reduction
steps within IRAF include bias subtraction, removal of pixel-to-pixel
sensitivity variations, optimal spectral extraction, and wavelength 
calibration based on helium arc-lamp spectra.
The peak signal-to-noise ratio of the individual spectra ranges from 50
to 125.
The RVs were measured using the lines H$\beta$, H$\gamma$, H$\delta$, H8 
and H9 through a cross-correlation analysis in which we used as a template a 
synthetic fit to an orbit-corrected average (all spectra shifted to zero 
velocity before averaging).

Finally, 32 medium-resolution spectra were obtained with MagE@Magellan\,I 
at Las Campanas Observatory in Chile
in 3.5\,hours on September 17, 2017, with 300\,s exposure times, 1\,arcsec 
slit, R$\sim$4100 and a useful wavelength range of 3500-8100\,\AA.
The typical signal-to-noise ratios were between 80 and 110.
The spectra were reduced using the MagE pipeline 
\citep{Kelson_2000,Kelson_2003}.
The RVs were measured using Balmer lines and two He\,I lines at 4471 
and 5875\,\AA.

\begin{figure}
\centering
\includegraphics[width=8.8cm,angle=0]{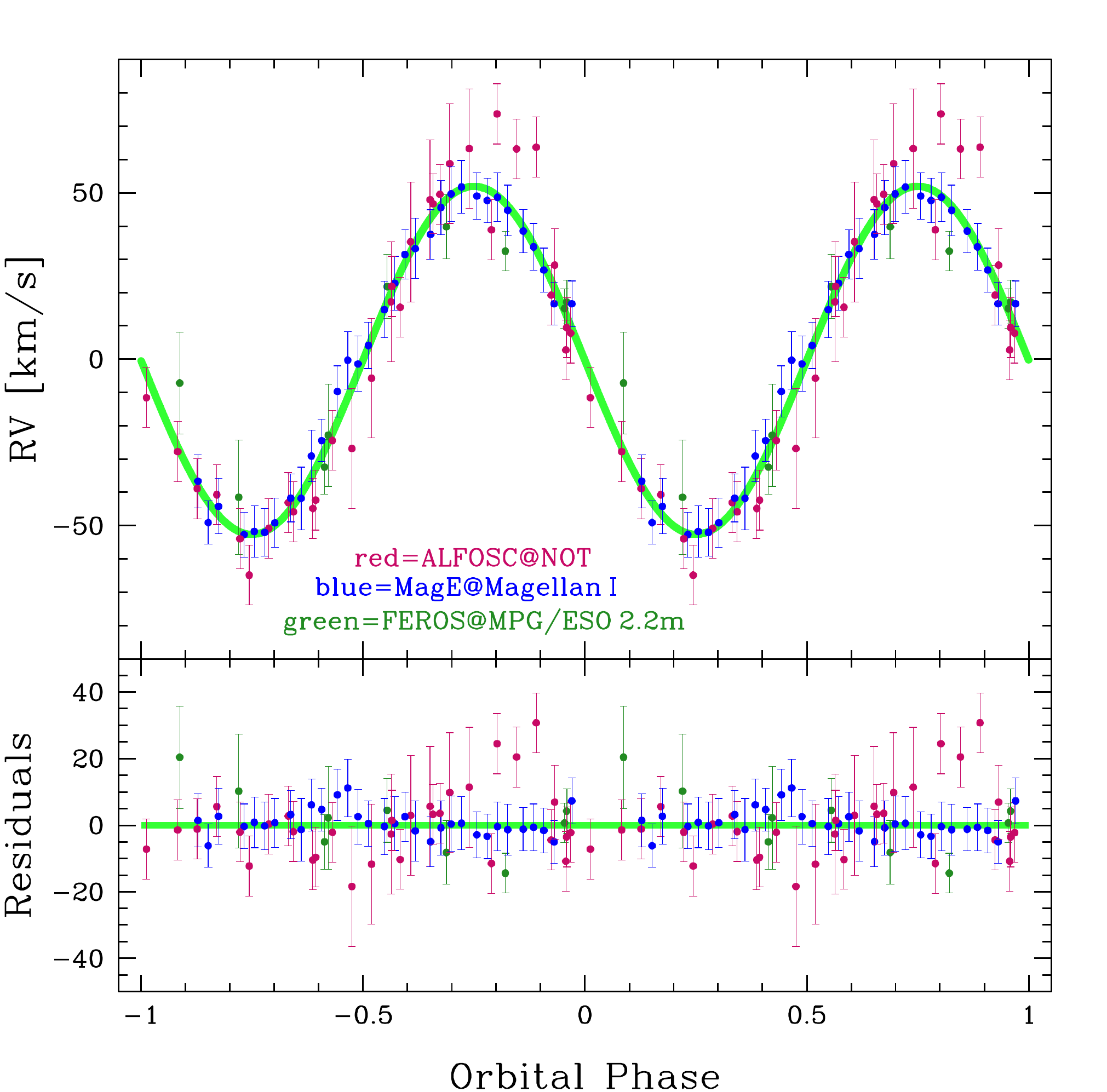}
\vspace{-3mm}
\caption{Radial velocities.}
\label{RVs}
\end{figure}

The RV measurements obtained from the MagE spectra are the most accurate 
due to the best compromise between high SN ratio, relatively high resolution 
and short exposure times, which means lower smearing.
However, the ALFOSC and FEROS RVs were also used using appropriate
weights (Fig.~\ref{RVs}).
From the best RV fit we obtain a circular orbit with a RV amplitude 
K=52.3$\pm$1.3~km/s, and a system velocity 
v$_0$=--6.4$\pm$1.2~km/s. Smearing is not considered as it is negligible
for MagE and ALFOSC spectra (0.08\% and 0.3\% respectively) and has little 
importance only for FEROS spectra (3\%).
By fitting all the 73 RVs listed in Table~\ref{RV_list} with an eccentric
solution, we can constrain the eccentricity to a value smaller than 
0.091\footnote{We obtain an eccentricity of 0.019$\pm$0.024, which 
translates into a 3$\sigma$ upper limit of 0.091.}.

Both the ALFOSC and MagE spectra were used not only to measure the radial 
velocities, but also to derive accurate atmospheric parameters and the 
rotational velocity of the sdOB star and to measure their variations as a 
function of the orbital phase, as described in the next section.

\begin{table} 
\centering
\caption[]{RV measurements.}
\begin{tabular}{lrrcc}
\hline\hline
~BJD$_{TDB}$ & \multicolumn{1}{c}{\hspace{-0.3mm} RV} & 
\multicolumn{1}{c}{\hspace{-1.0mm} error}  & instr. & 
UT date\\
\,\,\,--2450000.   & [km/s]                                     &
[km/s]                                         &                  &\\
\hline
7590.682242 &   11.14 &  6.64 & FEROS  & 2016-07-21\\ 
7590.755434 & --38.47 &  8.18 & FEROS  & 2016-07-21\\ 
7645.521343 & --28.90 & 15.42 & FEROS  & 2016-09-14\\ 
7645.542780 &   15.74 &  9.59 & FEROS  & 2016-09-14\\ 
7645.564206 &   33.73 &  9.60 & FEROS  & 2016-09-14\\ 
7645.585628 &   26.38 &  6.00 & FEROS  & 2016-09-14\\ 
7645.607066 &    9.17 &  6.00 & FEROS  & 2016-09-14\\ 
7645.628503 & --13.21 & 15.33 & FEROS  & 2016-09-14\\ 
7645.649944 & --47.56 & 17.09 & FEROS  & 2016-09-14\\ 
7958.472032 & --51.96 &  9.00 & ALFOSC & 2017-07-23\\
7958.479132 & --50.91 &  9.00 & ALFOSC & 2017-07-23\\
7958.486222 & --30.49 &  9.00 & ALFOSC & 2017-07-23\\
7958.493322 & --32.92 & 18.00 & ALFOSC & 2017-07-23\\
7958.500422 & --11.79 & 18.00 & ALFOSC & 2017-07-24\\
7958.507522 &   11.19 & 18.00 & ALFOSC & 2017-07-24\\
7958.514612 &   29.18 & 18.00 & ALFOSC & 2017-07-24\\
7958.521712 &   41.84 & 18.00 & ALFOSC & 2017-07-24\\
7958.528812 &   52.66 & 18.00 & ALFOSC & 2017-07-24\\
7958.535902 &   57.18 & 18.00 & ALFOSC & 2017-07-24\\
7958.565612 &   13.15 &  9.00 & ALFOSC & 2017-07-24\\
7958.572712 &    1.74 &  9.00 & ALFOSC & 2017-07-24\\
7958.579812 & --17.64 &  9.00 & ALFOSC & 2017-07-24\\
7958.591172 & --33.85 &  9.00 & ALFOSC & 2017-07-24\\
7958.598262 & --44.95 &  9.00 & ALFOSC & 2017-07-24\\
7958.605362 & --46.79 &  9.00 & ALFOSC & 2017-07-24\\
7960.478821 &   67.59 &  9.00 & ALFOSC & 2017-07-25\\
7960.485921 &   57.11 &  9.00 & ALFOSC & 2017-07-25\\
7960.493021 &   57.61 &  9.00 & ALFOSC & 2017-07-25\\
7960.549991 & --70.95 &  9.00 & ALFOSC & 2017-07-26\\
7960.557091 & --56.91 &  9.00 & ALFOSC & 2017-07-26\\
7960.564181 & --49.15 &  9.00 & ALFOSC & 2017-07-26\\
8014.490569 & --42.73 &  7.92 & MagE   & 2017-09-17\\
8014.494309 & --55.17 &  6.49 & MagE   & 2017-09-17\\
8014.498039 & --50.28 &  8.30 & MagE   & 2017-09-17\\
8014.507289 & --58.76 &  6.72 & MagE   & 2017-09-18\\
8014.511029 & --57.84 &  7.66 & MagE   & 2017-09-18\\
8014.514759 & --58.11 &  7.13 & MagE   & 2017-09-18\\
8014.518489 & --55.26 &  7.36 & MagE   & 2017-09-18\\
8014.524279 & --47.81 &  7.19 & MagE   & 2017-09-18\\
8014.528019 & --47.87 &  9.37 & MagE   & 2017-09-18\\
8014.531749 & --35.15 &  7.80 & MagE   & 2017-09-18\\  
8014.535489 & --30.53 &  6.37 & MagE   & 2017-09-18\\
8014.541189 & --15.77 &  7.71 & MagE   & 2017-09-18\\
8014.544929 &  --6.39 &  8.64 & MagE   & 2017-09-18\\
8014.548659 &  --7.49 &  8.26 & MagE   & 2017-09-18\\
8014.552399 &  --1.96 &  6.91 & MagE   & 2017-09-18\\
8014.558219 &    8.82 &  8.43 & MagE   & 2017-09-18\\
8014.561949 &   16.75 &  8.10 & MagE   & 2017-09-18\\
8014.565679 &   25.39 &  7.44 & MagE   & 2017-09-18\\
8014.569419 &   27.14 &  9.01 & MagE   & 2017-09-18\\
8014.574989 &   31.38 &  7.49 & MagE   & 2017-09-18\\  
8014.578729 &   39.47 &  8.14 & MagE   & 2017-09-18\\
8014.582469 &   43.56 &  8.29 & MagE   & 2017-09-18\\
8014.586199 &   45.64 &  7.96 & MagE   & 2017-09-18\\
8014.591779 &   42.95 &  6.97 & MagE   & 2017-09-18\\
8014.595519 &   41.63 &  6.68 & MagE   & 2017-09-18\\
8014.599249 &   42.56 &  7.36 & MagE   & 2017-09-18\\
8014.602999 &   38.63 &  7.62 & MagE   & 2017-09-18\\
8014.608619 &   32.41 &  6.52 & MagE   & 2017-09-18\\
8014.612359 &   27.67 &  7.08 & MagE   & 2017-09-18\\
8014.616089 &   20.70 &  6.67 & MagE   & 2017-09-18\\  
8014.619829 &   10.60 &  6.42 & MagE   & 2017-09-18\\
\hline  
\end{tabular}	    
\label{RV_list}	    
\end{table}	    
		    
\begin{table} 	    
\centering	    
\contcaption{}	    
\begin{tabular}{lrrcc}
\hline\hline	    
~BJD$_{TDB}$ & \multicolumn{1}{c}{\hspace{-0.3mm} RV} & 
\multicolumn{1}{c}{\hspace{-1.0mm} error}  & instr. & 
UT date\\
\,\,\,--2450000.   & [km/s]                                     &
[km/s]                                         &                  &\\
\hline
8014.626199 &   10.55 &  6.89 & MagE   & 2017-09-18\\
8201.763003 &   32.78 &  9.00 & ALFOSC & 2018-03-24\\
8211.685742 & --48.44 &  9.00 & ALFOSC & 2018-04-03\\
8268.635281 &    3.39 &  9.00 & ALFOSC & 2018-05-30\\
8269.702361 &    9.53 &  9.00 & ALFOSC & 2018-05-31\\
8304.505900 &   40.62 &  9.00 & ALFOSC & 2018-07-05\\
8304.596910 & --60.01 &  9.00 & ALFOSC & 2018-07-05\\
8307.568760 &   43.45 &  9.00 & ALFOSC & 2018-07-08\\
8312.607699 &  --3.29 &  9.00 & ALFOSC & 2018-07-13\\
8338.477179 &   15.78 &  9.00 & ALFOSC & 2018-08-07\\
8338.536299 &   22.17 & 11.10 & ALFOSC & 2018-08-08\\
\hline
\end{tabular}
\end{table}

\section{Atmospheric parameters and rotational velocity of the primary}

The reflection effect adds additional light to the sdOB spectrum, which varies
with phase. Because we can not model this contribution, each individual 
spectrum is matched separately to a grid of synthetic models to derive the 
effective temperature, gravity and helium abundance. If the contribution to 
the spectrum of the primary is significant, the resulting atmospheric 
parameters should show trends with orbital phase as a consequence of the 
varying light pollution. 
Indeed, such apparent variations of atmospheric parameters have been
found in other reflection binaries such as HW\,Vir 
\citep{1999MNRAS.305..820W},
HS\,0705+6700 \citep{2001A&A...379..893D} and most distinctively in 
HS\,2333+3937 \citep{Heber_2004}. 
The best estimate of the atmospheric parameters comes from data taken during 
secondary eclipse and just before and after primary eclipse, when the light 
pollution should be lowest.  

We closely follow the analysis strategy outlined by \citet{Heber_2004}.
The Balmer and helium lines in the ALFOSC and MagE spectra are used to 
determine effective temperature, gravity and helium abundance, and the 
projected rotation velocity v\,$\sin i$. 
Because the spectral resolution of the
ALFOSC spectra is insufficient for v\,$\sin i$ to be determined, the latter is
derived from the MagE spectra. The ALFOSC spectra show the entire Balmer 
series with a well defined continuum and can, therefore, be used to determine 
T$_\text{eff}$, $\log\,g$ and $\log\,y=\log\,(N(He)/N(H))$.  
%
%
For the MagE spectra, their wavy run of the continuum prohibits the Balmer 
lines to be used. However, they are very useful to derive the projected 
rotation velocity and allow us to investigate the helium ionisation 
equilibrium including lines not covered by the ALFOSC spectral range, from 
which an independent estimate of the effective temperature can be obtained. 
Since the 
helium lines are quite insensitive to gravity, the gravity had to be fixed in 
the analysis of the MagE spectra to $\log$ g = 5.56 derived from the ALFOSC 
spectra.     
We match the Balmer (H$\beta$ to  H\,11) and He {\sc I} (4471 and 4026 
\AA), as well as He\,{\sc II} 4686 and 4542 \AA\ line profiles in the ALFOSC 
spectra, and He {\sc I} (4471 and 5875 \AA) and He\,{\sc II} (4686 and 5411 
\AA) lines in the MagE spectra with a grid of synthetic spectra.

The models are computed using three codes. First, the {\sc Atlas12} code 
\citep{1996ASPC..108..160K} is used to compute the atmospheric structure 
(temperature/density stratification) in LTE. Non-LTE population numbers are 
then calculated with the {\sc Detail} code 
\citep{1981PhDT.......113G} and the coupled equations of 
radiative transfer and statistical equilibrium are solved numerically. In the 
final step, the {\sc Surface} code \citep{1981PhDT.......113G} 
computes the emergent spectrum based on the non-LTE occupation numbers provided
by {\sc Detail}. In this step detailed line-broadening tables are 
incorporated. All three codes have been updated recently 
\citep[see][]{2018A&A...615L...5I}. 
The impact of departures from LTE for hydrogen and helium on the 
atmospheric structure is modelled by feeding back population numbers calculated
by {\sc Detail} to {\sc Atlas12} and iterate. 
In addition, the occupation probability formalism \citep{1994A&A...282..151H} 
for hydrogen has been implemented and line broadening tables have been updated.
Stark broadening tables for hydrogen and neutral helium are taken from 
\citet{2009ApJ...696.1755T} and \citet{1997ApJS..108..559B}, respectively. 
The broadening of the lines of ionised helium was treated as described by 
\citet{1972ApJS...24..193A}.

\begin{figure}
\includegraphics[width=0.48\textwidth,angle=0]{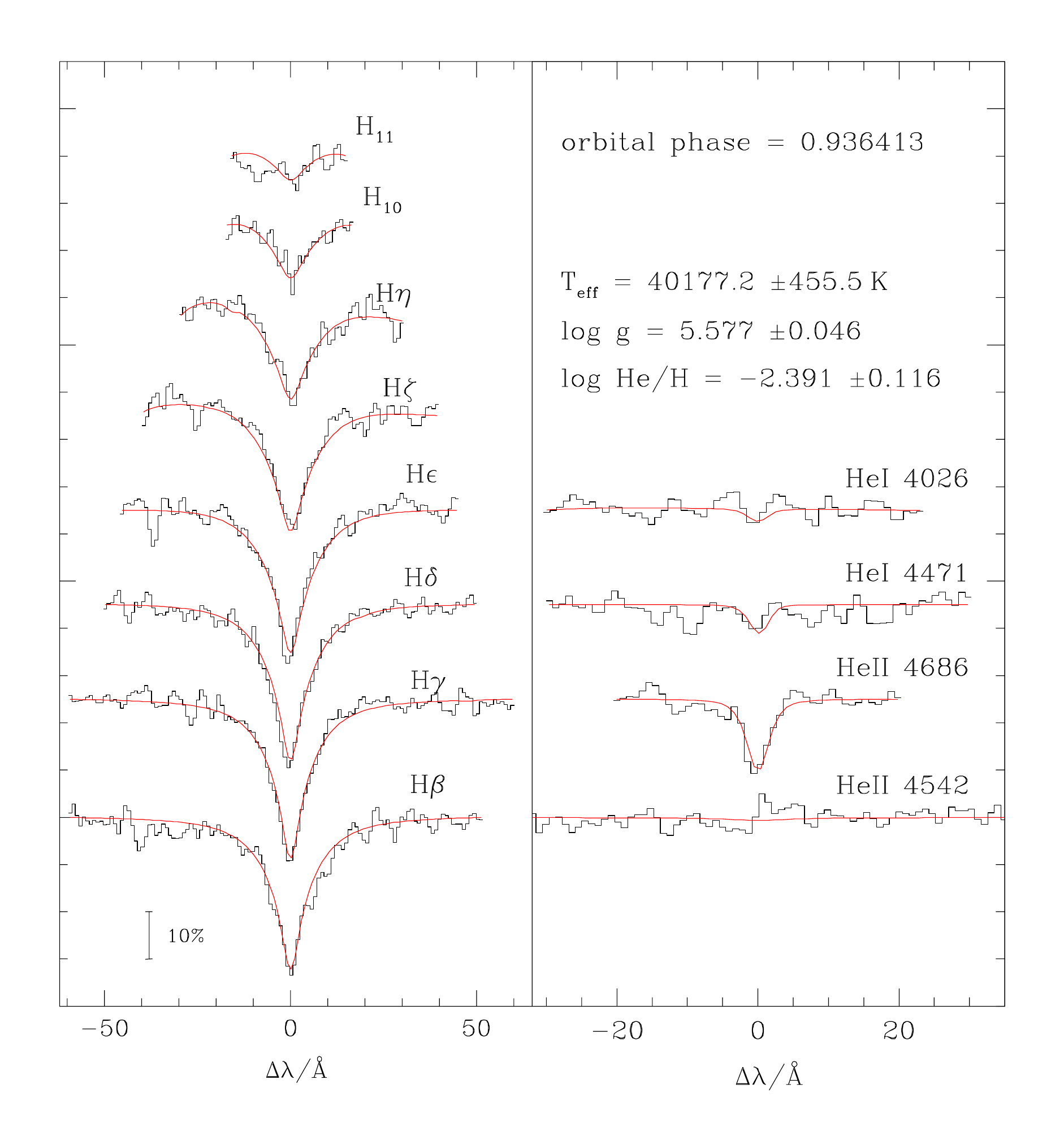}
\vspace{-5mm}
\caption{Fit of one of the ALFOSC spectra corresponding to orbital phase 
0.9364, close to the primary eclipse, for which the contribution of the 
secondary is minimum.}
\label{ALFOSC_fit}
\end{figure}

\begin{figure}
\includegraphics[width=0.48\textwidth,angle=0]{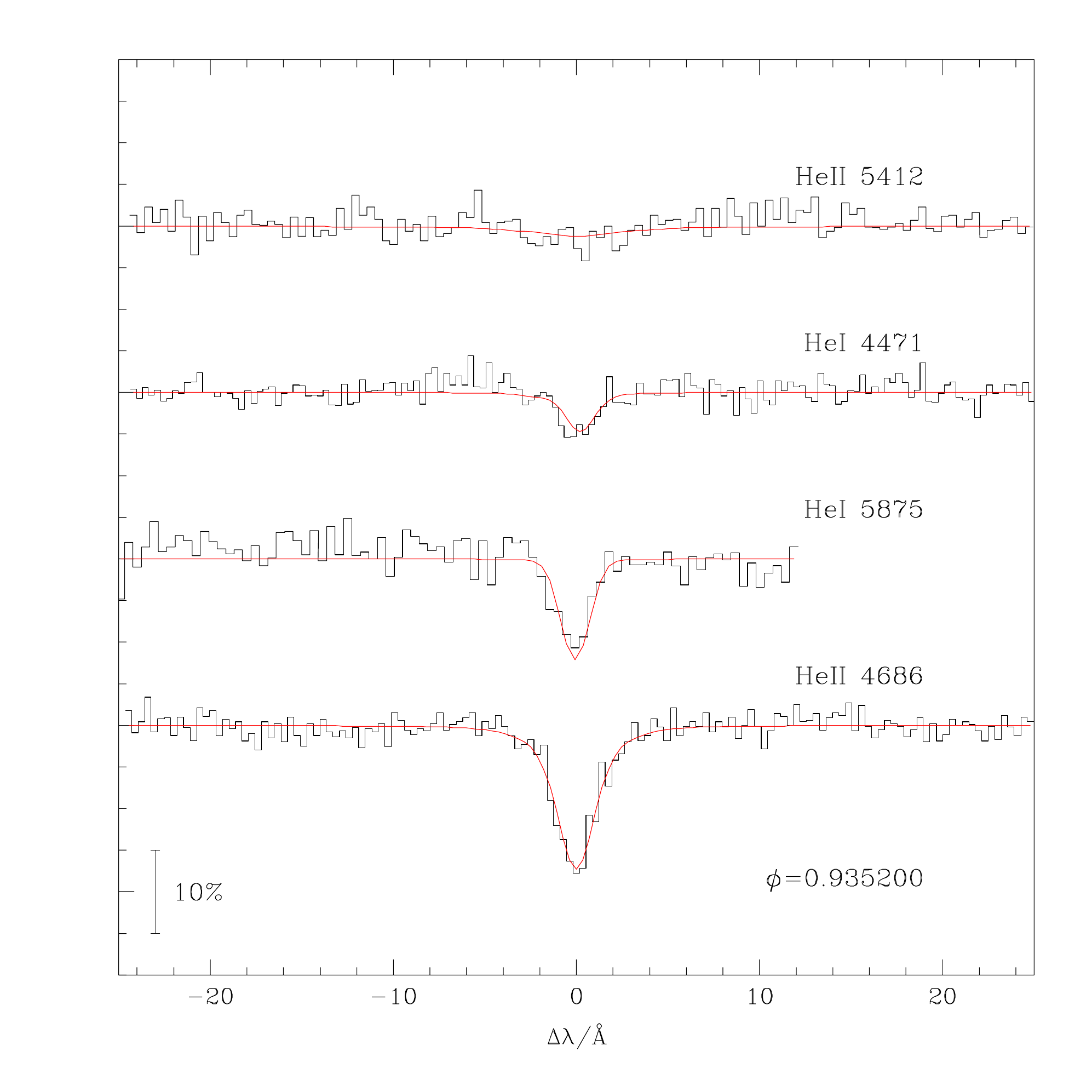}
\vspace{-5mm}
\caption{Same as Fig.~\ref{ALFOSC_fit} but for one of the MagE spectra 
corresponding to orbital phase 0.9352.}
\label{MagE_fit}
\end{figure}

The observed spectra are matched to the model grid by $\chi^2$ minimisation as 
described by \citet{1994ApJ...432..351S} using implementations by 
\citet{1999ApJ...517..399N} and \citet{2009PhDT.......273H}. Exemplary fits 
to an ALFOSC and a MagE spectrum are shown in Fig.~\ref{ALFOSC_fit} and
Fig.~\ref{MagE_fit}.

\begin{figure*}
\centering
\includegraphics[width=15.0cm,angle=0]{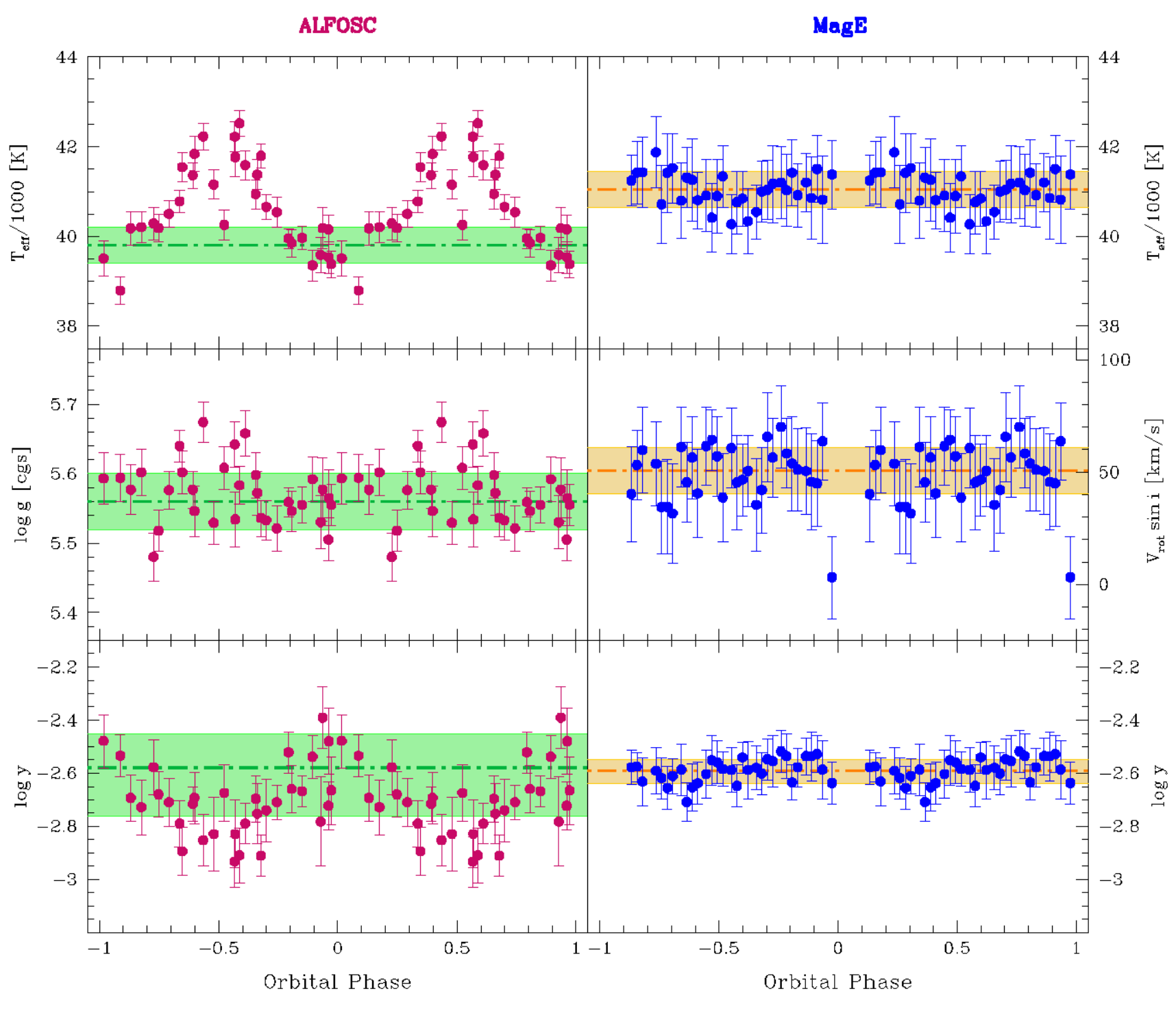}
\vspace{-5mm}
\caption{\teff, \logg, \logy\ and v$_{rot}$\,sin\,$i$ variations as a function 
of the orbital phase. Left panels: ALFOSC. The green dotted-dashed horizontal 
lines mark the adopted \teff, \logg, and \logy\ and the associated errors 
(cf. Table~\ref{Teffetc} and text). Note the two points with low \teff\ near 
phase 0.5 (secondary eclipse), when the contribution of the secondary is 
strongly reduced.
Right panels: MagE. The orange dotted-dashed horizontal lines mark the average 
values and associated errors of \teff, v$_{rot}$\,sin$i$, and He abundance. 
In the central panel note the outlier near phase zero (primary eclipse).}
\label{Teff_vs_phase}
\end{figure*}

The results of the quantitative spectral analysis of all ALFOSC and MagE 
spectra are summarized in Fig.~\ref{Teff_vs_phase}.
In the left panels (ALFOSC), the apparent variations of the effective 
temperature with phase and an amplitude of $\sim$2500\,K are obvious. 
The lowest temperatures ($\sim$40\,kK) occur near primary eclipse 
and in the secondary eclipse, where the contribution by extra light should be 
minimal. 
Hence, the increase during other orbital phases is caused by reflected light 
and, therefore, not real. Similarly, variations of the helium abundance are 
observed. The apparent variations of the surface gravity, however, are small. 
The mean values adopted for \teff, \logg\ and \logy, summarized in 
Table~\ref{Teffetc}, are obtained
by selecting six spectra close to the primary eclipse with phase between 
--0.1 and +0.1 and adding one spectrum at phase 0.52.
The analysis of the MagE spectra also results in effective temperatures and 
helium abundances that seem to vary with orbital phase 
(cf. Fig.~\ref{Teff_vs_phase} right panels),
but with amplitudes less pronounced than those from the ALFOSC spectra.
For this reason, we use all the MagE spectra to compute mean values 
and standard deviations of \teff\ and \logy\ (cf. Table~\ref{Teffetc}).

The effective temperature and surface gravity of EPIC\,216747137 
(\teff\,=\,40400\,K and \logg\,=\,5.56) are very similar to the hot HW\,Vir 
systems AA\,Dor \citep{2011A&A...531L...7K} and V1828\,Aql (=\,NSVS\,14256825, 
\citealt{2012MNRAS.423..478A}).
EPIC\,216747137 also shares an underabundance of helium (\logy=--2.59) 
with the two others. 

The projected rotational velocity, as derived from the individual MagE spectra 
(central right panel of Fig.~\ref{Teff_vs_phase}), results in a mean 
v\,$\sin$\,i\,=\,51$\pm$10\,km\,s$^{-1}$, significantly less than 
$\sim$70\,km\,s$^{-1}$ expected for tidally locked rotation.
Mean rotational velocity of 51 km\,s$^{-1}$ and standard deviation 
of 10 km\,s$^{-1}$ are 
obtained excluding only a single outlier close to phase 0
(see central right panel of Fig.~\ref{Teff_vs_phase}).

\begin{table}
\caption{SdOB atmospheric parameters and rotational velocity}
\begin{tabular}{lccc}
\hline\hline
 & ALFOSC & MagE & Adopted\\
\hline
\teff\,(K)   & 39800$\pm$400            & 41000$\pm$400            & 40400$\pm$1000\\
\logg~(cgs)  & 5.56$\pm$0.04            &                          & 5.56$\pm$0.06\\
\logy        & --2.58$^{+0.13}_{-0.18}$ & --2.59$^{+0.04}_{-0.05}$ & --2.59$\pm$0.05\\
v\,$\sin$\,i (km/s)&                    & 51 $\pm$ 10              & 51 $\pm$ 10\\
\hline
\end{tabular}
\label{Teffetc}
\end{table}

%

\section{Stellar Parameters: Radius, mass, and luminosity}

The second data release of Gaia provided a precise (5\%) parallax measurement 
which allows the stellar parameters (radius, mass, and luminosity) to be 
derived from the atmospheric parameters, if the angular diameter were known. 
The latter can be derived from the spectral energy distribution (SED).    

\subsection{Angular diameter and interstellar reddening}

The angular diameter $\Theta$ is derived from the observed flux $f(\lambda)$ 
and the synthetic stellar surface flux via the relation 
$f(\lambda)=\Theta^2 F(\lambda)/4$, which means that $\Theta$ is just a 
scaling factor which shifts fluxes up and down.
Strictly speaking, the apparent magnitudes of the sdOB can be measured during 
secondary eclipses only, when the companion is completely eclipsed by the 
larger subdwarf, because of the contamination by light from the companion's 
heated hemisphere. Such data are not available. Nevertheless, many photometric 
measurements are available in different filter systems, covering the spectral 
range in the optical and infrared. However, those measurements are mostly 
averages of observations taken at multiple epochs and, therefore, may be 
subject to light pollution from the companion. 

The low Galactic latitude ($b$\,=\,--9.9$^\circ$) implies that interstellar 
reddening may be large. Therefore the angular diameter has to be determined 
along with the interstellar colour excess. 
The reddening law of \citet{2019ApJ...886..108F} and a synthetic flux 
distribution from the grid of model atmospheres described in Sect.~4 is 
matched to the observed magnitudes employing a $\chi^2$ based fitting routine 
(see \citealt{Heber_2018} for details). 
The final atmospheric parameters and their respective uncertainties derived 
from the quantitative spectral analysis (see Sect.~4) are used. 
Indeed, interstellar reddening is significant with 
E(B-V)\,=\,0.213$^{+0.010}_{-0.016}$\,mag (see Table~\ref{SED_Gaia}).
The latter is consistent with values from reddening maps of 
\citet{Schlegel_1998} and \citet{Schlafly_2011}: 
0.253\,mag and 0.217\,mag, respectively. 

\begin{figure}
\centering
\includegraphics[width=8.5cm,angle=0]{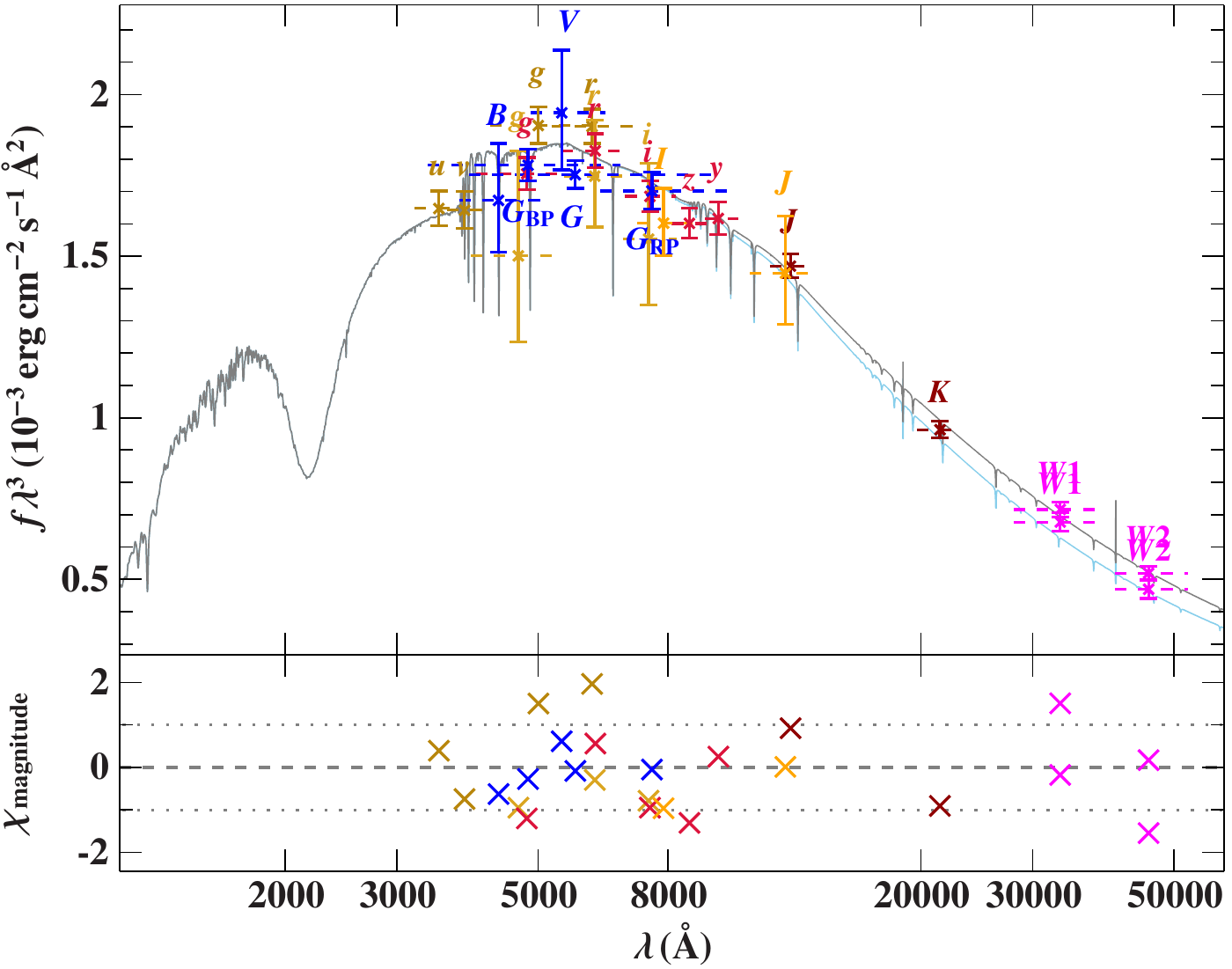}
\vspace{-3mm}
\caption{
Comparison of synthetic and observed photometry (flux times wavelength to the 
power of three): 
\textit{Top panel:} SED, filter-averaged fluxes converted from observed 
magnitudes. Dashed horizontal lines depict the approximate width of the 
respective filters (widths at tenth maximum). The  best-fitting model, smoothed
to a spectral resolution of 6\,\AA, is shown in gray.
\textit{Bottom panel:} residual $\chi$, difference between synthetic and 
observed magnitudes divided by the corresponding uncertainties.
The different photometric systems are assigned the following colours: SDSS 
(APASS, golden; \citealt{2015ApJS..219...12A});
SkyMapper (golden; \citealt{2018PASA...35...10W, 2019PASA...36...33O});
PAN-STARRS (red; \citealt{2017yCat.2349....0C});
Johnson (APASS, blue; \citealt{2015AAS...22533616H}); 
Gaia (blue; \citealt{2018A&A...616A...4E}
with corrections and calibrations from \citealt{2018A&A...619A.180M});
DENIS (yellow; \citealt{2000A&AS..141..313F});
VHS-DR6 (brown; \citealt{2007MNRAS.379.1599L});
WISE (magenta; \citealt{2014yCat.2328....0C, 2019ApJS..240...30S}).
}
\label{SED}
\end{figure}

Because of light pollution from the companion's heated hemisphere, the 
resulting angular diameter will be somewhat overestimated, as that is not
accounted for in the synthetic SED. Red and infrared magnitudes are expected to
be more affected than the blue ones. To account for the additional light, we 
added a black-body spectrum to the fit, allowing its temperature as well as the
relative emission area to vary. 
The final fit is shown in Fig.~\ref{SED} and results summarized in 
Table~\ref{SED_Gaia}. 

\begin{table}
\caption{SED\,+\,Gaia\,DR2 results}
\begin{tabular}{lr}
\hline\hline
\multicolumn{2}{l}{Atmospheric parameters from spectral analysis}\\
\hline	
Effective temperature \teff                     & 40400$\pm$1000 K\\
Surface gravity log$(g({\rm cm\,s^{-2}}))$      & 5.56$\pm$0.06\\
Helium abundance \logy                          & --2.59$\pm$0.05\\
\hline	
\multicolumn{2}{l}{Parameters from SED fit \& Gaia DR2 parallax}\\
\hline	
Color excess $E(B-V)$                           & 0.213$^{+0.010}_{-0.016}$ mag\\
Metallicity $z$ (fixed)                         & 0 dex\\
Angular diameter log($\Theta$(rad))             & --10.975$^{+0.009}_{-0.015}$\\
Black-body temperature $T_{bb}$                 & 2900$^{+2600}_{-1300}$ K\\
Black-body surface ratio $A_{eff,bb}$/$A_{eff}$ & 2.5$^{+4.3}_{-1.4}$\\
Generic excess noise $\delta_{excess}$ (fixed)  & 0.033 mag\\
Parallax $\varpi$ {\scriptsize (RUWE=1.04, offset=0.029 mas)$^*$} & 1.14$\pm$0.06 mas\\
\hline	
$R=\Theta/(2\varpi)$                            & 0.206$\pm$0.012 R$_{\odot}$\\
$M=gR^2/G$                                      & 0.56$^{+0.11}_{-0.10}$ M$_{\odot}$\\
$L/L_{\odot}=(R/R_{\odot})^2(T_{\rm eff}/T_{\rm eff,\odot})^4$ & 100$^{+16}_{-15}$\\
\hline
\multicolumn{2}{l}{\footnotesize $^*$ We use the RUWE parameter as a quality indicator,}\\
\multicolumn{2}{l}{\footnotesize \,\,\, best is 1, $<$1.4 is acceptable, 1.04 is good.}\\
\end{tabular}
\label{SED_Gaia}
\end{table}

\subsection{Stellar radius, mass and luminosity}

The Gaia DR2 parallax is corrected for a zero-point offset of $-$0.029\,mas as 
recommended by \citet{2018A&A...616A...2L} and applied by 
\citet{2018AJ....156...58B} to derive distances. 
By combining it with the atmospheric parameters (\logg\ and \teff) and the 
angular diameter, we can determine the star's radius R, mass M, 
and luminosity L. The respective uncertainties of the stellar 
parameters are derived by Monte Carlo error propagation. 
Results are summarized in Table~\ref{SED_Gaia}. 
Once the radius (R\,=\,0.206$\pm0.012$\,R$_\odot$) has been derived 
from angular diameter and parallax, the mass 
(0.56$^{+0.11}_{-0.10}$ M$_\odot$) follows from gravity and the luminosity 
(100$^{+16}_{-15}$ L$_\odot$) from radius and effective temperature.

\begin{figure}
\includegraphics[width=\linewidth]{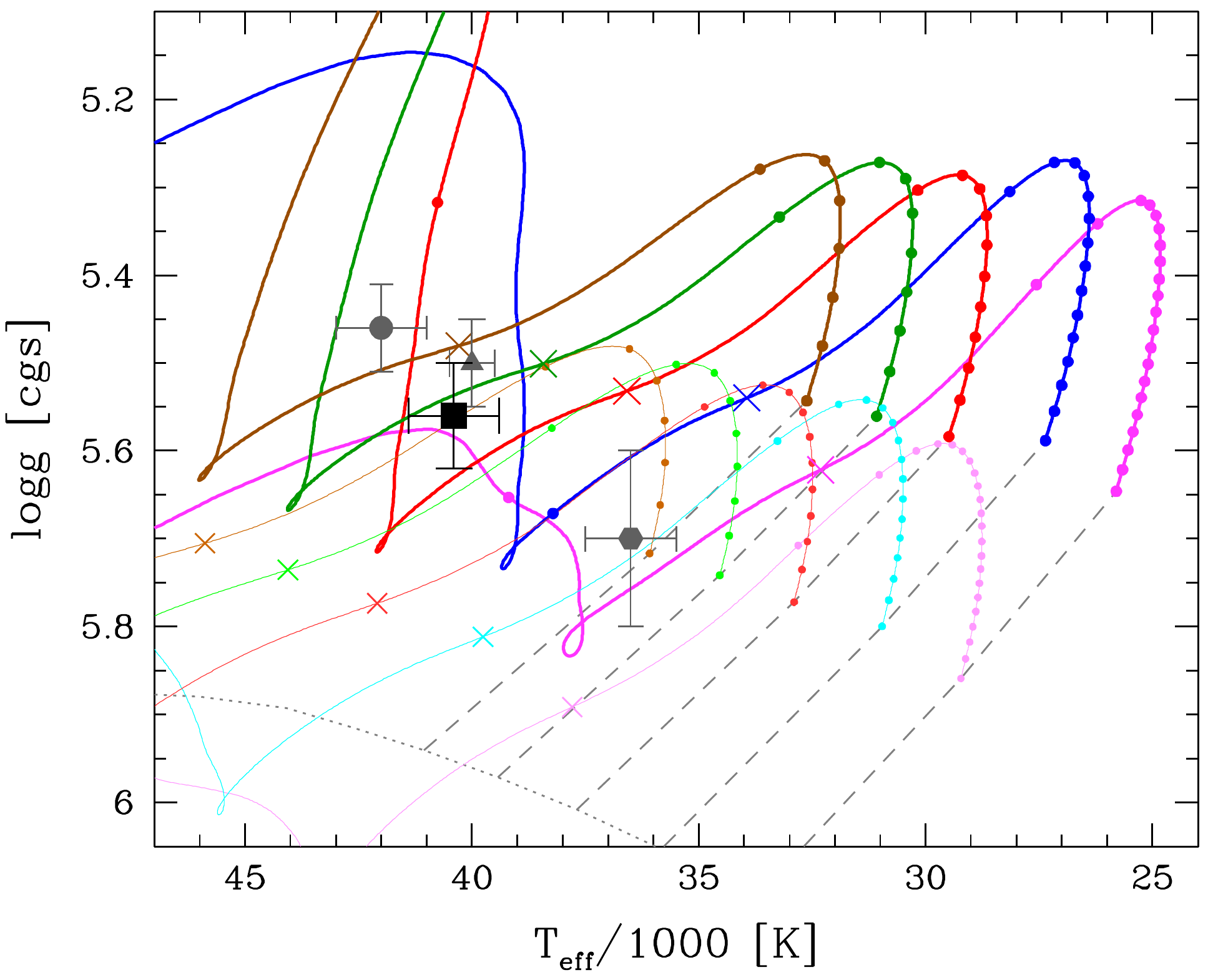}
\vspace{-3mm}
\caption{\teff-\logg\ diagram with the position of EPIC\,216747137 
(square symbol) compared with the evolutionary tracks by \citet{Han_2002}
for different stellar masses: 
from right to left 0.45, 0.50, 0.55, 0.60 and 0.65\,\msun, (magenta, blue, red,
green and brown respectively in the online version).
The envelope mass is 0.005\,\msun\ (thick lines) or 0.001 (thin lines and light
colours).
Along the evolutionary tracks, the age differences between adjacent
dots are 10$^7$\,years. The crosses mark the point of central helium 
exhaustion.
Helium main sequence and zero-age EHB (ZAEHB) are shown as dotted and
dashed lines respectively.
The positions of the evolved HW\,Vir systems
AA\,Dor (circle, \citealt{2011A&A...531L...7K}) 
and V1828\,Aql (triangle, \citealt{2012MNRAS.423..478A}) and of the
evolved reflection-effect sdB+dM binary HS\,2333+3927 (pentagon, 
\citealt{Heber_2004}) are also reported.
Note that AA\,Dor and V1828\,Aql have masses of 0.47 and 0.42 \msun\ 
respectively, and therefore should definitely be post-EHB (compare with the 
0.45\,\msun\ (magenta) and 0.50\,\msun\ (blue) tracks).}
\label{ev}
\end{figure}

A comparison with 
evolutionary models for EHB stars by \citet{Han_2002}
is shown in Fig.~\ref{ev} and demonstrates that the hot subdwarf has likely 
just evolved beyond the core-helium-burning phase, similar to AA\,Dor 
\citep{2011A&A...531L...7K} and V1828\,Aql \citep{2012MNRAS.423..478A}, or 
is at the very end of helium burning, depending on the hot subdwarf mass and 
envelope mass.

\section{Ephemeris}

First we computed independent ephemerides from photometric and RV data,
obtaining a good agreement on the orbital period.
The orbital period derived from the RVs has a higher precision thanks to the
longer baseline (2.0 vs 1.6 years) and also because of the poor time resolution
of the $K2$ data.
Then, considering both spectroscopic and photometric data together, we were
able to remove the degeneracy due to the spectral windows and obtain a better
determination of the orbital period thanks to the longer baseline (2.8 yrs).
%
%
In practice, taking as reference the center of the primary eclipse, 
we verified that the time difference between the last primary
eclipse of our data set (determined from RVs) and the first one (determined 
from photometry) was very close to an integer multiple of the orbital period 
determined from the RVs.
Then, imposing that such time difference is {\it exactly} a multiple of the 
orbital period, we obtain the best determination of the orbital period and 
the following best ephemeris:

\vspace{1mm}


\noindent
BJD$_{TDB}$=(2457301.56346\,$\pm$\,0.00041)+   

\hspace{7.8mm}
(0.16107224\,$\pm$\,0.00000017)\,E

\vspace{1mm}

\noindent
where BJD$_{TDB}$ is the barycentric Julian date of the center of each 
primary eclipse using barycentric dynamical time (see e.g. 
\citealt{Eastman_2010}).




\section{Modeling of the light curve}

The SAAO BRV light curves show relatively deep eclipses together with a 
reflection effect with increasing amplitude from B to R, and a secondary 
eclipse only visible due to the reflection effect. 
Such a light curve is characteristic for sdO/B systems with 
close, cool, low-mass companions. For the modeling of the light curve we used 
{\sc lcurve}, a code written to model detached but also accreting binaries 
containing a white dwarf \citep[for details, see][]{Copperwheat_2010}. 
It has been used to analyse several detached white dwarf-M dwarf binaries 
\citep[e.g.,][]{Parsons_2010}, which show very similar light curves with deep 
eclipses and a prominent reflection effect, if the primary is a hot white 
dwarf.
Recently, {\sc lcurve} was used also for an sdB+BD system
\citep[submitted]{Schaffenroth_2020}.
The code subdivides each star into small elements with a 
geometry fixed by its radius as measured along the direction towards the other 
star. 
Roche distortions and irradiation are also included, as well as limb-darkening,
gravitational darkening, lensing, Doppler beaming, R\o mer delay and 
asynchronous orbits. 
The latter three effects, lensing, Doppler beaming and R\o mer delay, are not 
detectable in our light curves. The irradiation is approximated by assigning a 
new temperature to the heated side of the companion:
%
%
%

\noindent
\hspace{2mm}
$\sigma T'^4_{\rm sec}=\sigma T^4_{\rm sec}+F_{\rm irr}=\sigma T^4_{\rm 
sec}\left[1+\alpha\left(\frac{T_{\rm prim}}{T_{\rm 
sec}}\right)^4\left(\frac{R_{\rm prim}}{a}\right)^2\right]$

\noindent
with $\alpha$ being the Bond albedo of the companion
%
%
and $F_{\rm irr}$ the irradiating flux, 
accounting for the angle of incidence and distance from the hot subdwarf. If 
the irradiation effect is very strong, the description given above might not be
sufficient. The backside of the irradiated star is completely unaffected in 
this description, but heat transport could heat it up, increasing the 
luminosity of unirradiated parts as well.

\begin{figure}
\includegraphics[width=\linewidth]{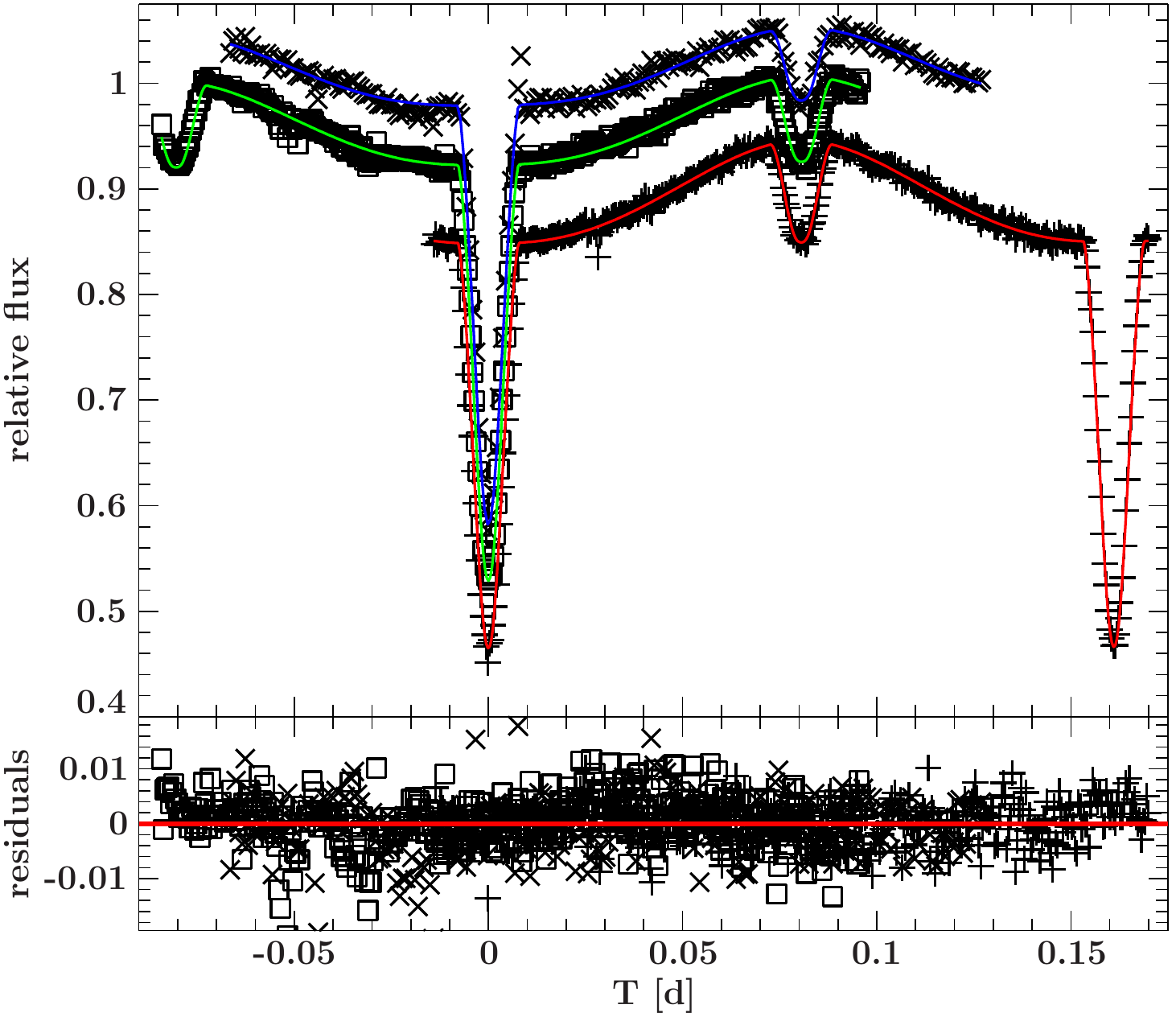}
\caption{Normalized SAAO B($\times$), V($\square$) and R(+)-band light curves 
together with the best fit. For better visualisation the V and R band light 
curves have been shifted. The lower panel shows the residuals.}
\label{lc}
\end{figure}

Since the model contains many parameters, not all of them independent, we 
fixed as many parameters as possible (see Table~\ref{param}).
The sdOB temperature was fixed to the temperature determined from the 
spectroscopic fit. 
The gravitational 
darkening coefficients were fixed to the values expected
for a radiative atmosphere for the primary 
\citep{von_Zeipel_1924}
%
%
and a convective atmosphere for the secondary \citep{Lucy_1967}, using a 
black-body approximation to calculate the resulting intensities.
More sophisticated models such as those proposed by \citet{2011A&A...533A..43E}
or \citet[see also \citealt{2020A&A...634A..93C}]{Claret_2011} were not used 
because the deformations from a spherical shape are very small and in fact
gravity darkening has almost no impact.
For the limb darkening of the primary, we adopted a quadratic limb darkening 
law using the tables by \citet{Claret_2011}. 
As the tables include only surface gravities up to $\log{g}=5$, we used the 
values closest to the parameters derived by the spectroscopic analysis. 
As the two stars are almost spherical (we do not see 
significant ellipsoidal deformations), the light curve is not sensitive to the 
mass ratio and therefore we computed solutions with different, fixed mass 
ratios. To localize the best set of parameters, we used a SIMPLEX algorithm 
\citep{press92} varying the inclination, the radii, the temperature 
of the companion, the geometric albedo of the companion ($A_2$), 
the limb darkening of the 
companion, the period and the time of the primary eclipse. Moreover, we also 
allowed for corrections of a linear trend, which is often absorbed in observing
hot stars, as the comparison stars are often redder and so the correction for 
the air mass is often insufficient (slope). The model of the best fit is shown 
in Fig.~\ref{lc}, together with the observations and the residuals.

To get an idea about the degeneracy of the light curve solutions, as well as 
the errors of the parameters, we performed also Markov-Chain Monte-Carlo (MCMC)
computations using the best solution obtained with the SIMPLEX algorithm as a 
starting value and varying the radii, the inclination, the temperature of the 
companion, as well as the albedo of the companion 
(Fig.~\ref{mcmc_b},\ref{mcmc_v},\ref{mcmc_r}). 
A clear correlation between the radius of the companion, the inclination, and 
the geometric albedo of the companion ($A_2$) can be seen, which results from 
the fact that the companion is only visible in the light curve due to the 
reflection effect and the amplitude depends on the inclination, the radius of 
the companion and the albedo, as well as the separation and temperature and 
radius of the primary, which is given by the spectroscopic analysis.

\begin{table}\caption{Parameters of the light curve fit of the SAAO BVR-band 
light curves for the best model.}
\begin{tabular}{llll}
\hline
\hline
band & B & V & R\\
\hline
\multicolumn{4}{c}{Fixed Parameters}\\
\hline
q                 & \multicolumn{3}{c}{0.175}\\
$P$               & \multicolumn{3}{c}{0.1610732}\\
$T_{\rm eff,sdB}$ & \multicolumn{3}{c}{40400}\\
$x_{1,1}$ & 0.0469 & 0.0434 & 0.0379\\
$x_{1,2}$ & 0.2668 & 0.2346 & 0.2082\\
$g_1$             & \multicolumn{3}{c}{0.25}\\
$g_2$             & \multicolumn{3}{c}{0.08}\\
\hline
\multicolumn{4}{c}{Fitted parameters}\\
\hline
$i$ & $85.04\pm0.40$ & $85.62\pm0.19$ & $85.51\pm0.14$\\
$r_1/a$ & $0.1887\pm0.0016$ \hspace{-5mm} & $0.1890\pm0.0008$ \hspace{-5mm} & $0.1887\pm0.0005$\\
$r_2/a$ & $0.1251\pm0.0028$ \hspace{-5mm} & $0.1216\pm0.0012$ \hspace{-5mm} & $0.1222\pm0.0009$\\
$T_{\rm eff,comp}$ \hspace{-5mm} & $3000\pm500$ & $2965\pm482$ & $3042\pm503$\\
$A_2$ & $0.95\pm0.08$ & $1.01\pm0.04$ & $1.25\pm0.04$\\
$x_2$ & 0.33 & 0.27 & 0.28\\
$T_0$ & 2457892.53884 & 2457891.57235 & 2457893.66629\\
slope & 0.004729 & 0.003015 & 0.00088\\
$\frac{L_1}{L_1+L_2}$ & 0.98028 & 0.972691 & 0.94864\\
\hline
\end{tabular}
\label{param}
\end{table}

\begin{table}\caption{Absolute parameters of the system}
\begin{tabular}{lcc}
\hline\hline
& ~~~~~~best solution & ~~~~post-EHB canonical\\
\hline
$q$                           & ~~~~~~0.175	      & ~~~~0.194	    \\
$a\,[\rm R_\odot]$            & ~~~~~~$1.121\pm0.028$ & ~~~~$1.028\pm0.025$ \\
$M_{\rm sdOB}\,[\rm M_\odot]$ & ~~~~~~$0.620\pm0.023$ & ~~~~$0.470\pm0.017$ \\
$M_{\rm comp}\,[\rm M_\odot]$ & ~~~~~~$0.109\pm0.004$ & ~~~~$0.091\pm 0.003$\\
$R_{\rm sdOB}\,[\rm R_\odot]$ & ~~~~~~$0.212\pm0.005$ & ~~~~$0.194\pm0.005$ \\
$R_{\rm comp}[\rm R_\odot]$   & ~~~~~~$0.137\pm0.003$ & ~~~~$0.125\pm0.003$ \\
$\log g_{\rm phot}$\,[cgs]    & ~~~~~~$5.58\pm0.01$   & ~~~~$5.54\pm0.01$   \\
\hline	
\end{tabular}
\label{results}
\end{table}

\section{Nature of the companion}

As stated before, it is not possible to derive the mass ratio from the 
light-curve analysis. 
Since we have only a single-lined system, it is necessary to 
look for other possibilities to constrain the mass ratio of the system. 
Taking into account the sdOB atmospheric parameters obtained from our 
spectroscopic analysis, the sdOB star is likely an evolved post-EHB star or 
just at the end of helium burning, depending on the hot subdwarf mass and 
envelope mass.

\begin{figure}
\includegraphics[width=1.04\linewidth]{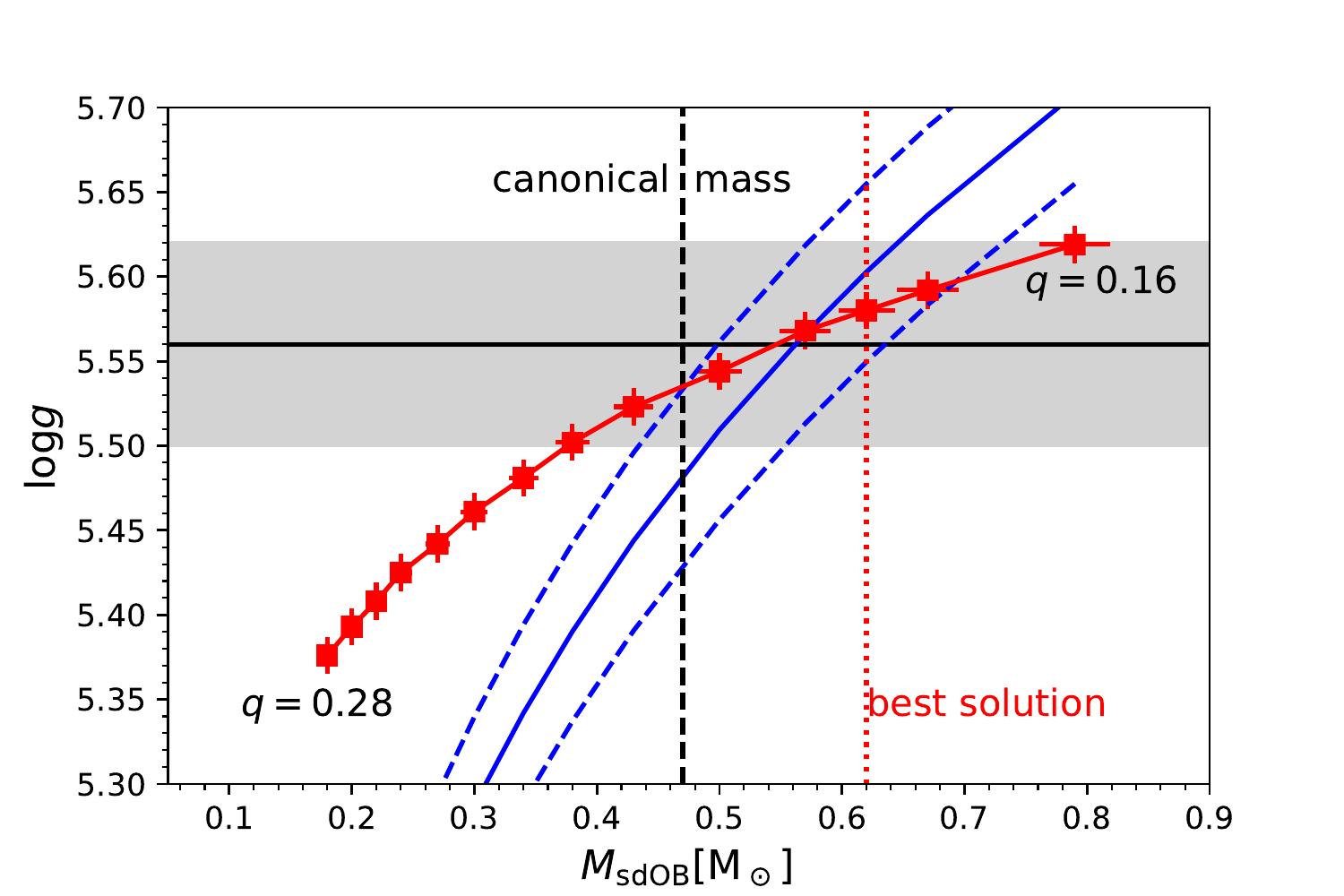}
\vspace{-3mm}
\caption{Mass of the sdOB versus the photometric \logg\ for different mass 
ratios from 0.16 to 0.28 in steps of 0.1 (adding 0.175 for the best solution). 
They were derived from combining the 
results from the analysis of the light curve and radial velocity curve. The 
grey area marks the spectroscopic \logg\ that was derived by the spectroscopic 
analysis. The blue lines mark the surface gravity derived from the radius 
determined by the Gaia parallax and the SED fit. The vertical lines represent 
the two solutions which are given in Table~\ref{results}.}
\label{sdb}
\end{figure}

When we combine the analysis of the radial velocity curve and the light curve, 
we get different masses and radii of both components, as well as a different 
separation for each solution with a different mass ratio. From the 
spectroscopic analysis we derived the surface gravity of the hot subdwarf, 
which can be compared to a photometric surface gravity calculated from the mass
and radius derived from the light-curve analysis and the mass function. 
Moreover, from the radius determined by the Gaia parallax and the SED fit, we
can calculate a Gaia surface gravity. The comparison of the photometric, 
spectroscopic, and Gaia surface gravity is shown in Fig.~\ref{sdb}. An 
agreement is seen for a mass between $\sim$0.47 and $\sim$0.67 \msun. This 
means a post-EHB hot subdwarf with a canonical mass of 0.47 \msun\ cannot be 
excluded.

\begin{figure}
\includegraphics[width=1.02\linewidth]{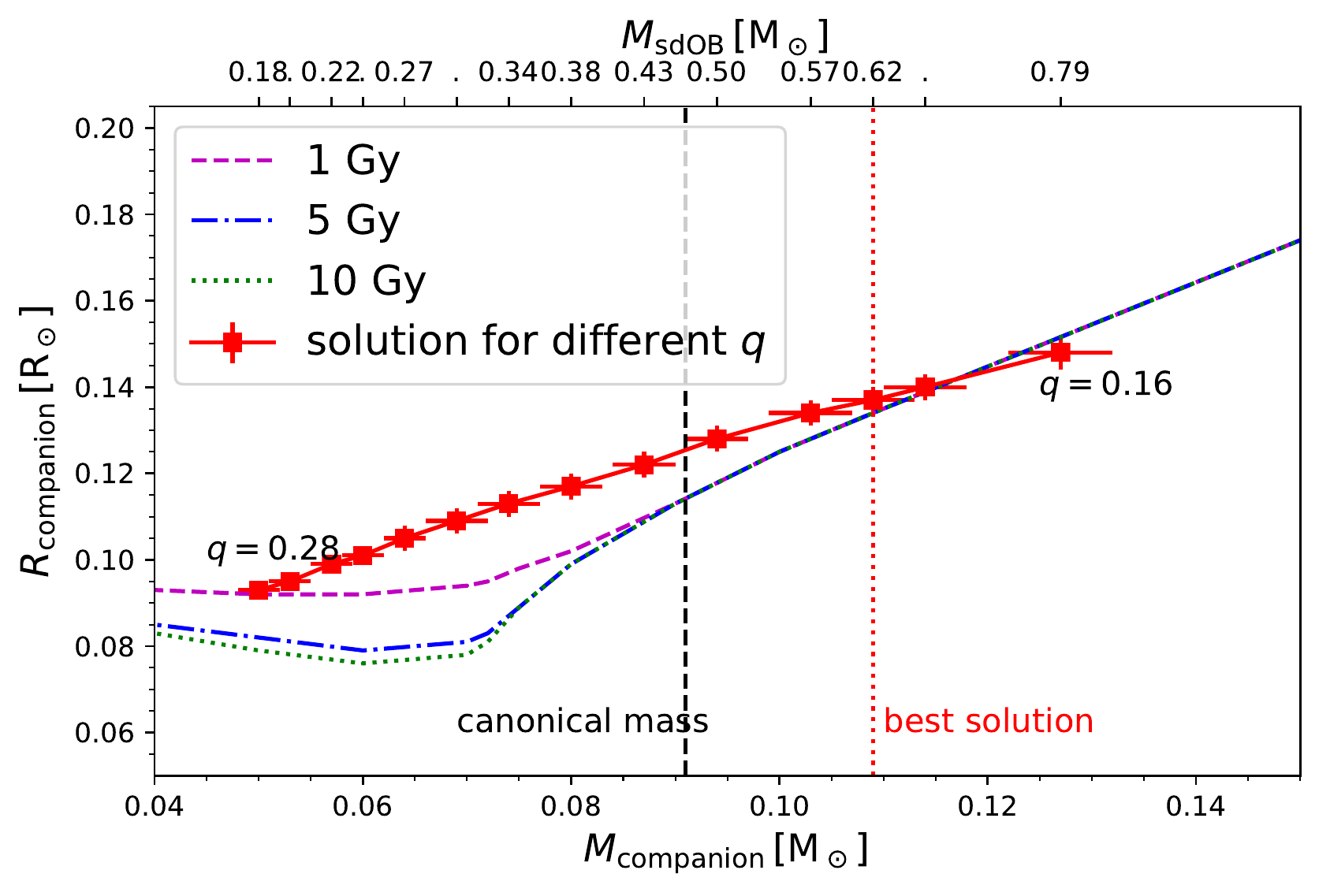}
\vspace{-4mm}
\caption{Comparison of theoretical mass-radius relations of brown dwarfs by 
\citet{Baraffe_2003} and low-mass M dwarfs by \citet{Chabrier_1997} for an age 
of 1 Gyr (dashed), 5 Gyr (dotted-dashed) and 10 Gyr (dotted) to results from 
the light cure analysis. Each error cross represents a solution from the light 
curve analysis for a different mass ratio ($q = 0.16-0.28$ in steps of 0.1 and
adding 0.175 for the best solution). 
The vertical lines represent the two solutions of Table~\ref{results}.}
\label{bd}
\end{figure}

Another possibility to constrain the masses further is to consider the 
mass-radius relation of the companion (Fig.~\ref{bd}), and compare it to 
theoretical predictions \citep{Baraffe_2003,Chabrier_1997}. Using the 
mass-radius relation for the cool companion, the best agreement is found for a 
sdOB mass of $\sim0.62\,\rm M_\odot$. This is hence the most consistent 
solution, that implies a 
$\sim$2\%
inflation of the M-dwarf radius. 
A lower mass would imply a more inflated radius for the M dwarf.

In Table~\ref{results} we consider two solutions (absolute system parameters)
of the light curve analysis resulting from two different assumptions on the 
mass ratio q. A massive one at q=0.175, corresponding to an sdOB mass of 0.62
M$_\odot$, which we prefer because it avoids strong inflation of the companion,
and a second solution at q=0.194, which corresponds to the canonical mass 
(M=0.47 M$_\odot$). 
For the preferred solution of a high mass post-EHB star, we obtain a companion 
mass of $0.109\pm0.004\,M_\odot$, corresponding to a low-mass M-dwarf. 
For a canonical mass sdOB, the mass of the M star would be even less 
($0.091\pm0.003\,M_\odot$), only slightly above the stellar mass limit.

\section{Summary and discussion}

EPIC\,216747137 is a new HW\,Vir system that belongs to the small subgroup of
eclipsing hot subdwarf binaries in which the primary is a hot, evolved, sdOB 
star.
The other two members of this group, AA\,Dor and V1828\,Aql, with a mass
of 0.47 and 0.42 \msun\ respectively 
\citep{2011A&A...531L...7K,2012MNRAS.423..478A}, should definitely be 
post-EHB stars (and this is particularly true for AA\,Dor that has been 
intensively studied by various teams).
While for EPIC\,216747137, due to its larger mass of $\sim$0.62, we can just 
say that it is close, and likely beyond, central helium exhaustion.

Among the 20 published HW\,Vir systems, only AA\,Dor, V1828\,Aql and
EPIC\,216747137 have effective temperatures near 40 kK,
while all the others have \teff\ between 25 and 35 kK, compatible with He-core 
burning \citep{2018OAst...27...80W}.
Moreover, these three hotter HW\,Vir systems seem to
follow a different relation in the \teff-\logy\ plane 
\citep{2003A&A...400..939E} respect to all the other HW\,Vir stars.
The position of all the published HW\,Vir in a \teff-\logy\ plane can be seen 
in \citet[Fig.~5]{2018OAst...27...80W}.
Since the number of new HW\,Vir systems is rapidly increasing, with 25 
new systems already spectroscopically confirmed and many more to come 
\citep{Schaffenroth_2019},
the larger statistics will allow us to confirm or not
that HW\,Vir stars follow two different sequences in the \teff-\logy\ plane.

The orbital period of EPIC\,216747137, $\sim$0.161 days, and the mass of its 
dM companion, $\sim$0.11 \msun, fit well with the period distribution and the 
companion mass distribution of the hot subdwarf binaries with a dM companion 
\citep[Fig.~7 and 8]{Kupfer_2015}.
However, in the preferred light curve solution, the sdOB mass is unusually high
(0.62 \msun). Such a high mass could result from post-AGB evolution, but this
possibility is ruled out because it would imply a luminosity ten times higher 
than observed.
When we consider constraints from spectroscopy, light curve solution and 
parallax, the mass must be between 0.47 and 0.67 \msun.
Hence a mass as low as 0.47 \msun\ can not be ruled out, 
but it implies that the cool companion is significantly inflated.
Although inflation in M dwarfs is not a well understood phenomenon
\citep[see e.g.][]{2018MNRAS.481.1083P}, a strong inflation 
appears quite unlikely, and this is why we prefer the high-mass option.

A mass as high as $\sim$0.62 \msun\ provides a challenge for the hot subdwarf 
formation theories since the CE ejection channel struggles to form stars with 
a mass higher than $\sim$0.47, while the RLOF channel does not work for 
orbital periods shorter than $\sim$1 day 
(see e.g. \citealt[Fig.~12 and 10 respectively]{Han_2003}).

Another interesting aspect of our results is that EPIC\,216747137 is not 
synchronized.
Among the other 9 systems with published rotational velocities, only three of 
them are not synchronised 
\citep[submitted, and references therein]{Schaffenroth_2020}, all of them being
relatively young and not evolved (and with a BD 
candidate companion, but this might be related to a selection effect 
considering that it is easier to obtain high-resolution data when the 
companion is a BD), while the other six more evolved systems are all 
synchronised.
The growing number of synchronized systems seems in contradiction with the 
prediction by \citet{2018MNRAS.481..715P} that synchronization time-scales
are longer than the sdB lifetime.

Hot subdwarf stars are found in all stellar populations 
\citep{2017MNRAS.467...68M,2020ApJ...898...64L}. 
EPIC\,216747137 lies just 155\,pc below the Galactic plane. 
This hints at thin disc membership. In order to check this assumption, 
we carried out a kinematical investigation calculating Galactic trajectories 
in a Galactic potential (for details see appendix B). 
The Galactic orbit is almost perfectly circular and the binary orbits within 
(though close to) the solar circle (Fig.~\ref{galactic_orbit}).
Hence, we conclude that the binary belongs to the thin disc population, 
which is also confirmed by its position in the Toomre diagram 
(Fig.~\ref{toomre}). 


\section*{Acknowledgements}

The $K2$ data presented in this paper was obtained from the Mikulski Archive 
for Space Telescopes (MAST). Space Telescope Science Institute is operated by 
the Association of Universities for Research in Astronomy, Inc., under NASA 
contract NAS5-26555.
This paper uses observations made at the South African Astronomical
Observatory (SAAO).
The spectroscopic results are based on observations collected 
at the MPG/ESO 2.2\,m telescope; 
%
%
at the 2.6\,m Nordic Optical Telescope (NOT), operated jointly by Denmark, 
Finland, Iceland, Norway, and Sweden;
and at the 6.5\,m Magellan\,I telescope.
RS acknowledges financial support from the INAF project on
``Stellar evolution and asteroseismology in the context of the PLATO space 
mission'' (PI S. Cassisi).
VS is supported by the Deutsche Forschungsgemeinschaft, DFG through grant 
GE 2506/9-1.
DK thanks the University of the Western Cape and the National Research
Foundation of South Africa for financial support.
We thank Jantje Freudenthal for support at the SAAO 1\,m Elizabeth telescope,
Michel Rieutord and an anonymous referee for useful comments.
RS wishes to remember and thank Nicol\`o D'Amico, President of INAF,
who suddenly passed away a few days before the submission of this article.
%


\section*{Data availability}

The data underlying this article will be shared on reasonable request to the 
corresponding author.



\bibliographystyle{mnras.bst}
\bibliography{EPIC_216_EB_FINAL_arXiv} 








\appendix

\section{MCMC fits of the SAAO BVR light curves}

\onecolumn

\begin{figure}
\includegraphics[width=\linewidth]{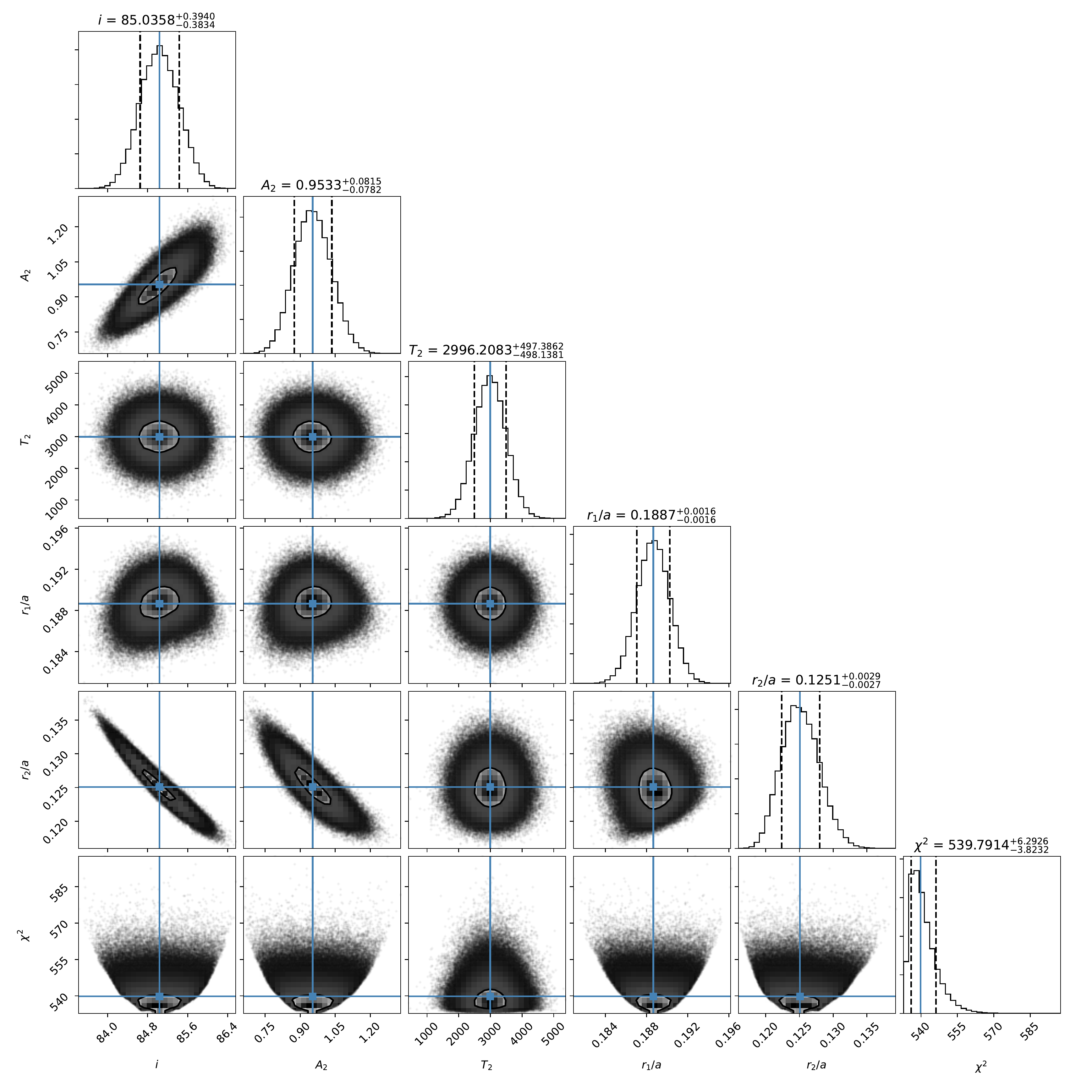}
\caption{MCMC computations showing the degeneracy and the parameter 
errors of the B-band light curve solutions.}
\label{mcmc_b}
\end{figure}



\begin{figure}
\includegraphics[width=\linewidth]{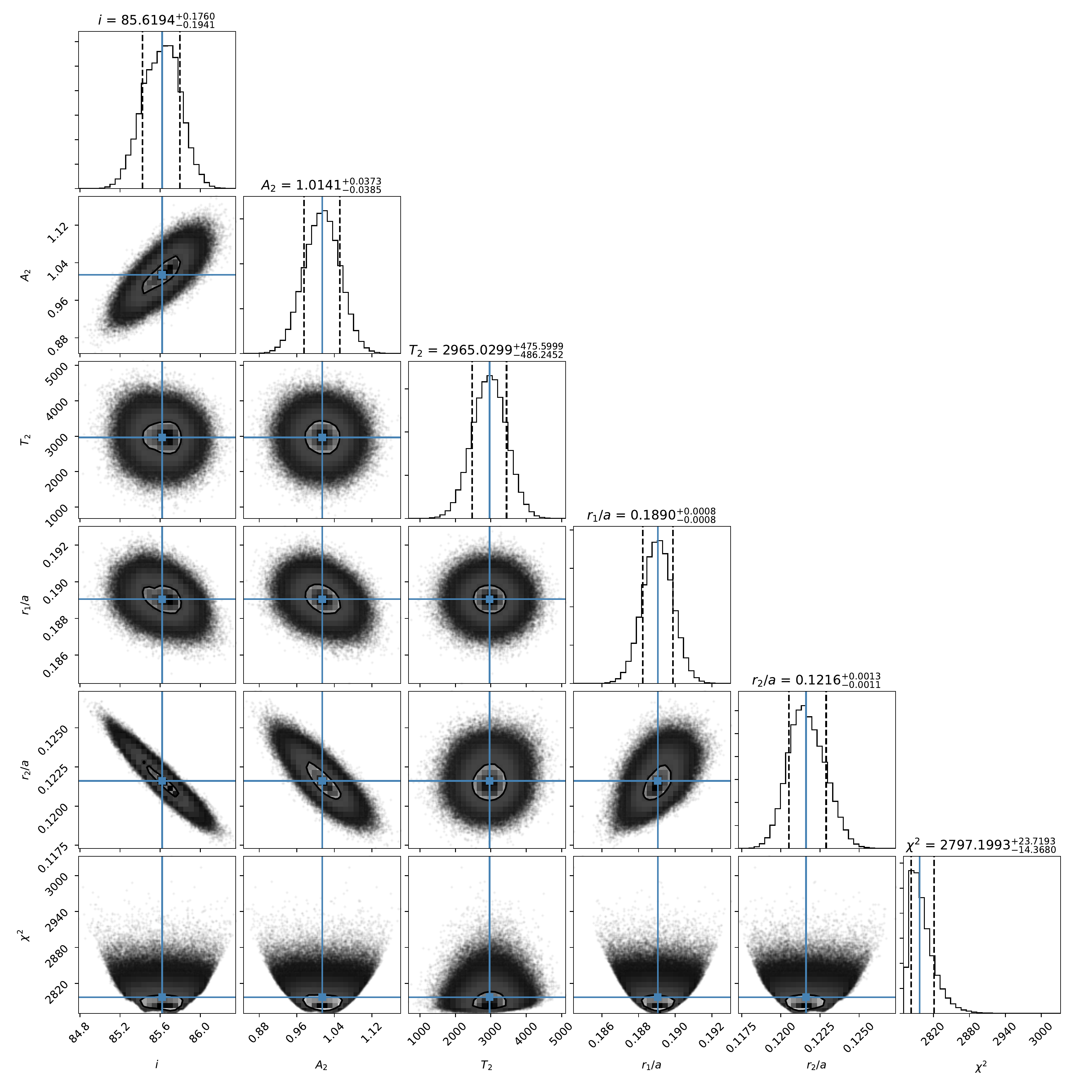}
\caption{Same as Fig.~\ref{mcmc_b} but for the V-band light curve.}
\label{mcmc_v}
\end{figure}



\begin{figure}
\includegraphics[width=\linewidth]{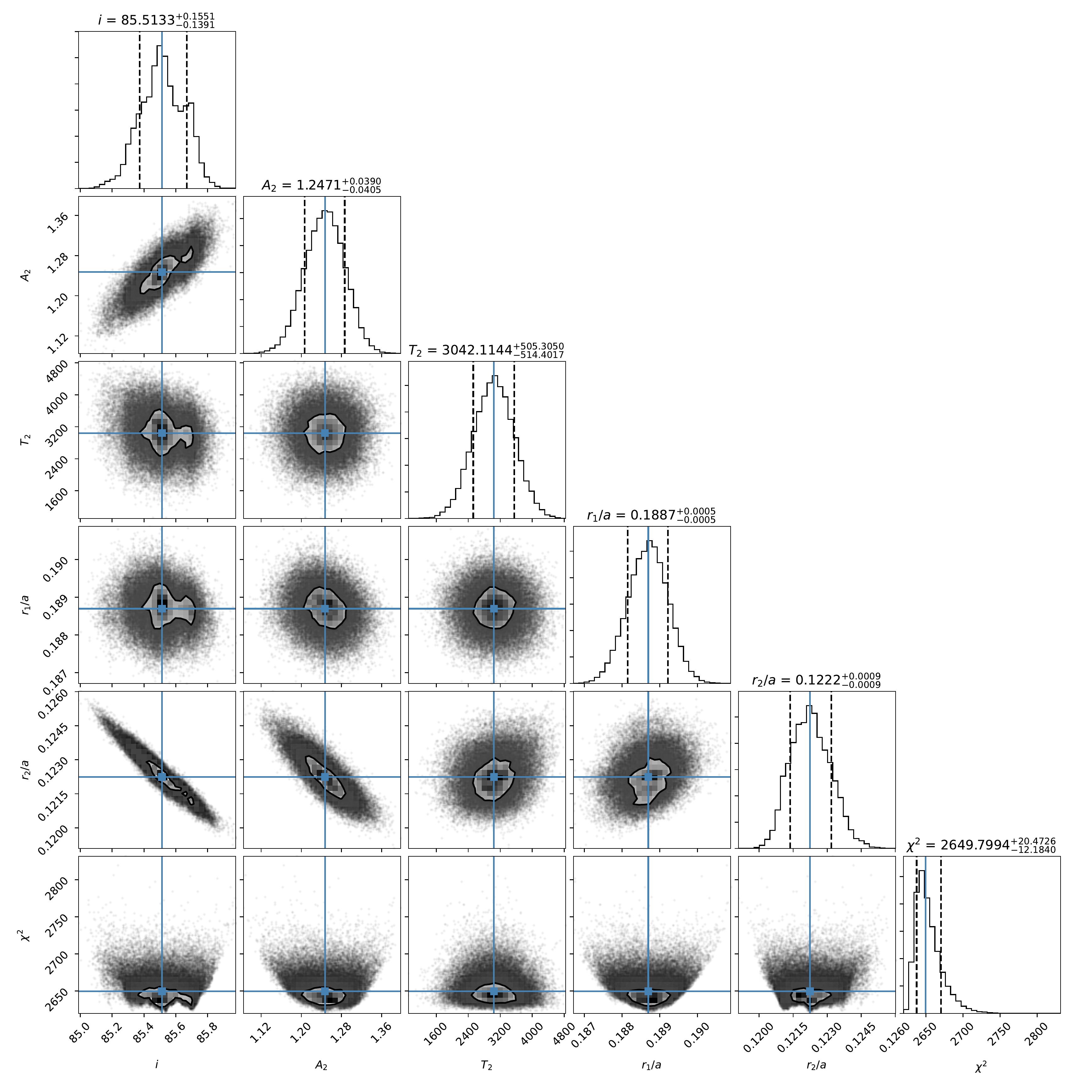}
\caption{Same as Fig.~\ref{mcmc_b} but for the R-band light curve.}
\label{mcmc_r}
\end{figure}


\twocolumn
\newpage

\vspace{16mm}

\section{Kinematics of EPIC\,216747137}

\vspace{16mm}

\begin{figure}
\includegraphics[width=\linewidth]{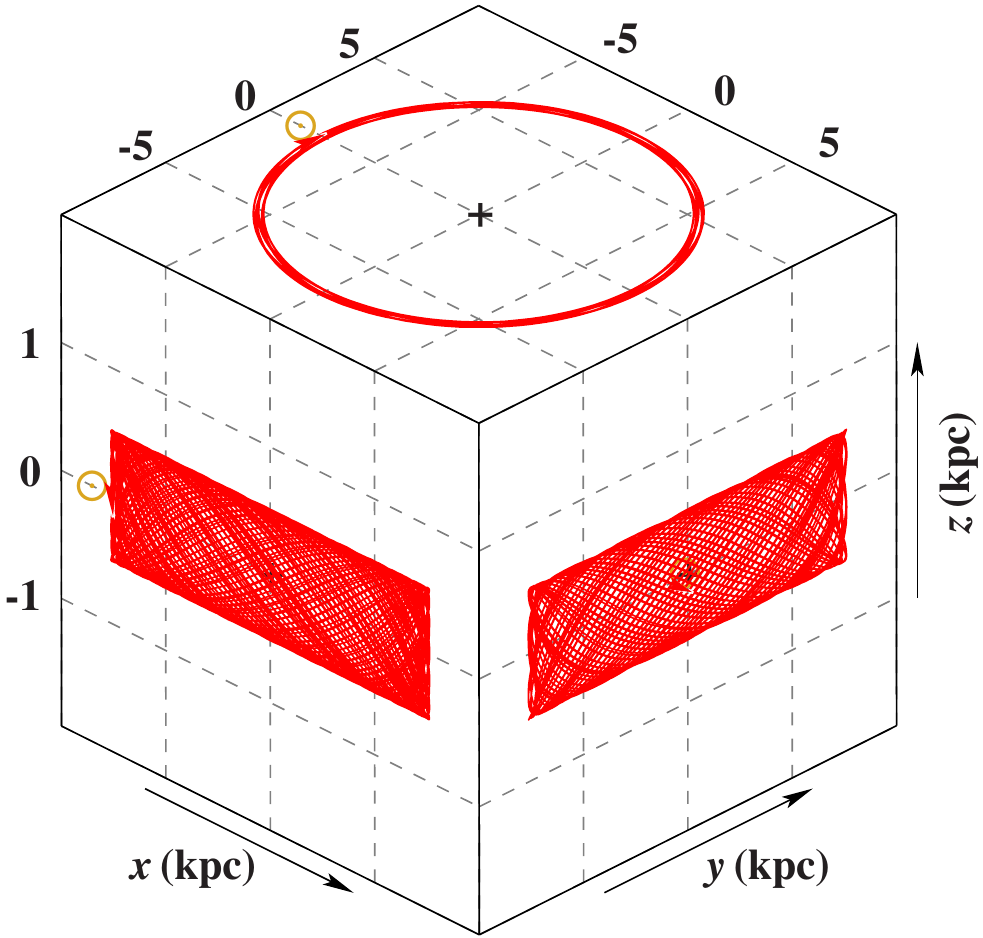}
\caption{EPIC\,216747137's three-dimensional orbit in a Cartesian Galactic 
coordinate system. 
The centre of the Galaxy lies at the origin, the Sun (yellow circled dot) on 
the negative x-axis. The z-axis points to the Galactic north pole. 
Trajectories were computed back in time for 10 Gyrs using a standard, 
axisymmetric model for the Galactic gravitational potential 
\citep[an updated version of that of \citet{1991RMxAA..22..255A}, see][for 
details]{2013A&A...549A.137I}. 
The shape of the orbit is almost circular, with vertical oscillations of a few 
hundred pc amplitude, typical for a thin-disk star 
(see e.g. \citealt{2006A&A...447..173P}).}
\label{galactic_orbit}
\end{figure}

\begin{figure}
\includegraphics[width=\linewidth]{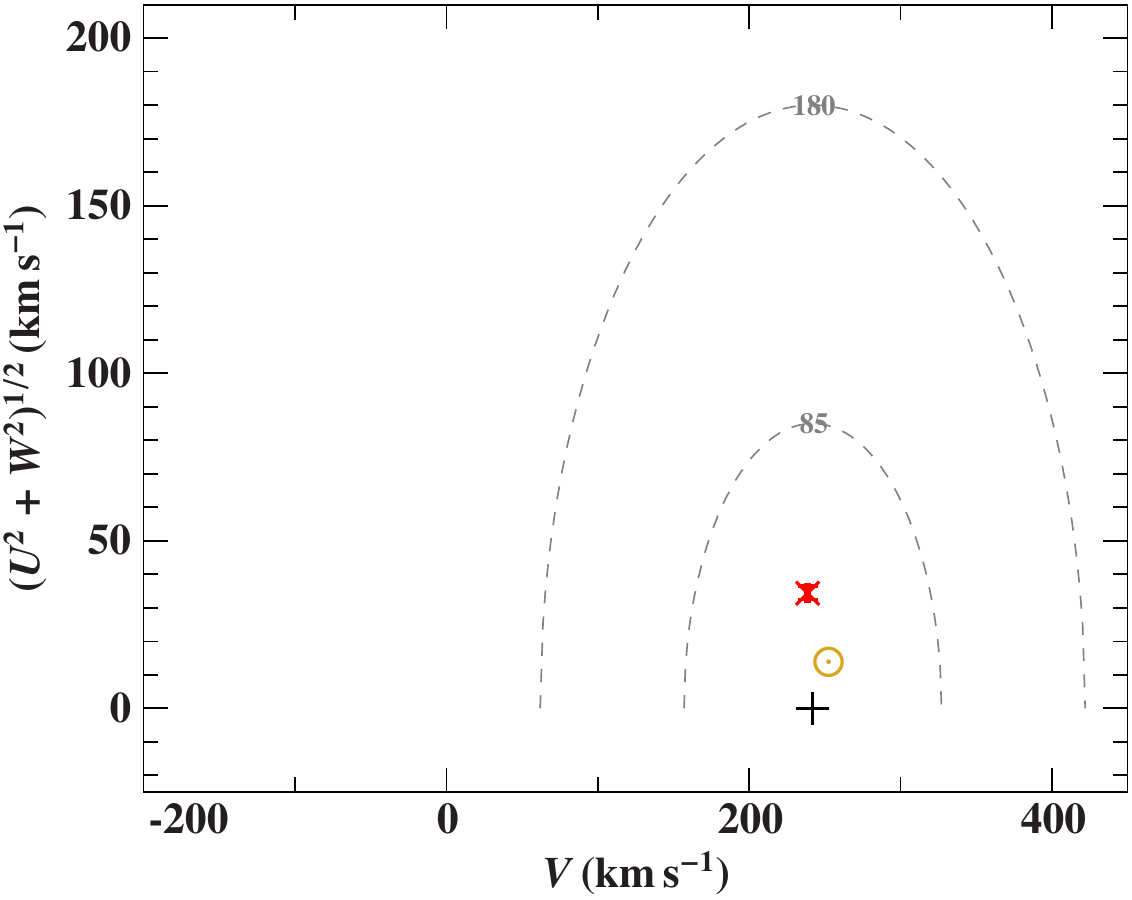}
\caption{The position of EPIC\,216747137 (red cross with $1\sigma$ error bars) 
in the Toomre diagram. 
The velocity component $V$ is measured in the direction of the rotation of the
Galaxy, $U$ towards the Galactic centre, and $W$ perpendicular to the plane. 
The yellow circled dot marks the position of the Sun. The local standard of 
rest (LSR) is marked by a plus sign. According to 
\citet{2004AN....325....3F}, the boundaries for thin and thick disk are 
located at $85$\,km\,s${}^{-1}$ and $180$\,km\,s${}^{-1}$ respectively 
(dashed circles centered around the LSR).}
\label{toomre}
\end{figure}



\bsp	
\label{lastpage}
\end{document}

%% file: defs.tex
\newcommand{\teff}{\ensuremath{T_{\mathrm{eff}}}}
\newcommand{\logg}{\ensuremath{\log g}}
\newcommand{\logy}{\ensuremath{\log y}}

\newcommand{\msun}{${\mathrm{M}}_{\odot}$}

\newcommand{\mjup}{$\rm M_{\rm Jup}$}

\def\mnras{MNRAS}
\def\aj{AJ}%
\def\apj{ApJ}%
\def\apjl{ApJ}%
\def\aap{A\&A}%
\def\pasp{PASP}%
\def\apss{Ap\&SS}
\def\nat{Nature}